\documentclass[use AMS,use natbib]{mn2e}
\usepackage{amsmath}
\usepackage{amssymb}
\usepackage[T1]{fontenc}
\usepackage[latin1]{inputenc}
\usepackage{framed} 
\usepackage{color}
\usepackage{multirow}
\usepackage{textcomp}
\usepackage{units}
\usepackage{subfigure} 

\usepackage{tabularx}
\usepackage{array}
\usepackage{color}

\usepackage[]{epic,eepic,latexsym,epsf}
\usepackage[xdvi]{graphicx}

\begin{document}
\title{Magnetic field amplification and X-ray emission in galaxy minor mergers}
\author[A. Geng et al.]{Annette Geng$^{1}$\thanks{E-mail: annette.geng@uni-konstanz.de}, Hanna Kotarba$^{2}$, Florian B\"{u}rzle$^{1}$, Klaus Dolag$^{2,3}$, \newauthor Federico Stasyszyn$^{2}$, Alexander Beck$^{2}$ and Peter Nielaba$^{1}$ \newauthor\\
$^{1}$University of Konstanz, Department of Physics, Universitätsstr. 10, 78464 Konstanz, Germany\\
$^{2}$University Observatory Munich, Scheinerstr. 1, 81679 Munich, Germany\\
$^{3}$Max Planck Institute for Astrophysics, Karl-Schwarzschild-Str. 1, 85741 Garching, Germany} 

\maketitle

\begin{abstract}
We investigate the magnetic field evolution in a series of galaxy minor mergers using the N-body/smoothed particle hydrodynamics (SPH) code \textsc{Gadget}. The simulations include the effects of radiative cooling, star formation and supernova feedback. Magnetohydrodynamics (MHD) is implemented using the SPH method. We present 32 simulations of binary mergers of disc galaxies with mass ratios of 2:1 up to 100:1, whereby we have additionally varied the initial magnetic field strengths, disc orientations and resolutions. We investigate the amplification of a given initial magnetic field within the galaxies and an ambient intergalactic medium (IGM) during the interaction. We find that the magnetic field strengths of merger remnants with mass ratios up to 10:1 saturate at a common value of several $\mu$G. For higher mass ratios, the field strength saturates at lower values. The saturation values correspond to the equipartition value of magnetic and turbulent energy density. The initial magnetization, disc orientation and numerical resolution show only minor effects on the saturation value of the magnetic field. We demonstrate that a higher impact energy of the progenitor galaxies leads to a more efficient magnetic field amplification. The magnetic and turbulent energy densities are higher for larger companion galaxies, consistent with the higher impact energy supplied to the system. We present a detailed study of the evolution of the temperature and the bolometric X-ray luminosity within the merging systems. Thereby we find that magnetic fields cause a more efficient increase of the IGM temperature and the corresponding IGM X-ray luminosity after the first encounter. However, the presence of magnetic fields does not enhance the total X-ray luminosity. Generally, the final value of the X-ray luminosity is even clearly lower for higher initial magnetic fields. 
\end{abstract}


\section[]{Introduction}

In the framework of hierarchical growth of structure in the universe, galaxy interactions are believed to be an 
essential part of galaxy formation and evolution. The Lambda cold dark matter ($\Lambda$CDM) cosmology predicts 
the formation of dark matter haloes due to gravitational instabilities in the
early universe which later on form larger
 haloes by gas accretion and halo mergers (\citealt{WhRe78,WhFr91}). In the continuing process, baryonic 
particles get gravitationally bound to the haloes, forming structures which are consistent with the galaxies observed today. The merger rate of dark matter haloes is an increasing function of redshift (\citealt{KoBu99,GoKl01}), implying that collisions and mergers were much more frequent in the early
Universe. Moreover, simulations of merger history trees \citep{WeBu02} indicate that major merger events were comparatively rare, and thus dark matter haloes grow mainly by the accretion of smaller objects. More precisely, minor mergers are expected to be at least one order of magnitude more common than major mergers (\citealt{HeMi95}, and references therein). 

The most obvious consequence of galaxy collisions, the shaping of the participating objects, was first described by \citet{ToTo72}. 
Interactions of galaxies lead to significant
changes of the dynamics of the progenitor galaxies due to the alteration of the gravitational potential (e.g. \citealt{ToTo72, NaBu03}). Thereby, most of the galaxy collisions result in a merger of the progenitor systems.

So far, simulations of galaxy mergers were predominantly dedicated to studies of star 
formation, stellar dynamics, gas flows, supermassive black holes or feedback from stars and black holes (e.g. \citealt{CoJo08,DiSp05,SpDi05a,SpDi05b,RoBu06,JoNa09}).
However, galaxy mergers are also interesting in the context of the amplification and restructuring of
 small-scale magnetic fields within the scope of the global evolution of cosmic magnetism \citep[e.g.][]{KoLe09,KoKa10}. 
Observations of galaxies have revealed that virtually all galaxies host magnetic fields, with field strengths between 1 and 10  $\mu$G (see e.g. \citealt {BeBr96} for a review; \citealt{ChBo07,VoSo10}).

For comparison, the intergalactic magnetic field is usually estimated to be less than $10^{-8}$ G \citep[e.g.][]{KrBe08}. The comparatively strong galactic magnetic fields are commonly explained by the action of a galactic dynamo \citep[for a review see e.g.][and references therein]{Ku99}. This dynamo, which is derived using the mean field theory, acts on the basis of the conversion of poloidal into toroidal fields and vice versa. Thereby, turbulent motions generated by stellar activity are transformed into cyclonic motions due to Coriolis forces. The cyclonic motions are responsible for the conversion of toroidal flux into poloidal flux ($\alpha$-effect). 
The differential rotation of the galactic disc causes the conversion of poloidal into toroidal flux ($\Omega$-effect). This $\alpha\Omega$ dynamo leads to an efficient amplification of the magnetic field. 
A similar approach exploits the Parker instability \citep{Park92} caused by cosmic ray (CR) particles to generate large-scale turbulence which in turn results in an efficient dynamo action. This CR-driven dynamo process has been recently simulated by \citet{HaOt09}, showing an equipartition of the turbulent energy with magnetic field energy and cosmic ray energy within the gas. 
However, despite their success in explaining the magnetic field strengths and structures in fully evolved local galaxies, these dynamos are challenged by recent observations of strong ($\mu$G) magnetic fields at high redshifts \citep{BeMi08}. Therefore, further amplification processes of galactic magnetic fields acting on short timescales have to be investigated. \citet{ArBe09} used analytical considerations to show the importance of turbulence in protogalactic haloes, which may lead to efficient small-scale dynamo action. This work thus already showed the importance of turbulence for the evolution of magnetic fields in the early universe. However, the conclusions made by \citet{ArBe09} are based on only rough estimates of the amplification timescales and the turbulence driven by galactic collisions. 

In order to investigate the idea of an interaction-driven amplification of galactic magnetic fields, \citet{KoKa10} have performed numerical simulation of a galactic (major) merger of the Antennae Galaxies, including the evolution of magnetic fields. They found that the magnetic field within the colliding system gets amplified by compression and shear flows up to a saturation value of $\approx 10 \mu$G, independent of the initial magnetic field of the progenitor discs, which was varied between $10^{-9}$ and $10^{-6}$ G. Within this work, the saturation level was found to be near equipartition between magnetic and turbulent gas pressure, in good agreement with theoretical expectations of the turbulent dynamo \citep[e.g.][and references therein]{ArBe09}. In a continuative study, \citealt{KoLe11} considered a major collision of three galaxies. These studies confirmed the saturation of the galactic magnetic field at the equipartition level of several $\mu$G independent of the initial magnetic field. Furthermore, an additionally included ambient IGM allowed also for studies of its magnetization, which was shown to saturate at  $\approx 10^{-8}$G. However, all of these studies are dedicated to galactic major mergers. As galaxy minor mergers are expected to be far more frequent within the process of structure formation, it is definitely interesting to consider the influence of the mass ratios of the progenitor galaxies on the magnetic field amplification and saturation value. This idea is pursued within the presented work.

Moreover, \citealt{KoLe11} also found that the initial magnetization of the galaxies and the IGM affects the propagation of interaction-driven shocks within the IGM. Thereby, the shocks are stronger (gaining higher Mach numbers) for higher initial magnetizations. This is also indicated by a larger extent of the shock heated regions and the higher temperatures of these regions. This hot gas is expected to radiate in X-rays due to thermal bremsstrahlung. 

Observations with the \textit{ROSAT} X-ray telescope \citep{RePo98} and the \textit{Chandra} X-ray observatory (e.g. \citealt{FaZe01}) of interacting galaxy systems revealed the production of extended X-ray emission during the interactions. Simulations focusing on the X-ray emission of interacting galaxies were recently performed by \citet{CoDi06} and \citet{SiHo09}. \citet{CoDi06} considered the X-ray emission generated by interactions of galaxies including gaseous discs, but without including an ambient IGM gas. They found that the cold gas in the galactic discs gets shock-heated to X-ray emitting temperatures during the collision. The X-ray emission in their simulations (different major merger scenarios with different gas fractions) increases during the progress of the collision and peaks at the time of the final merger. They showed that galaxy mergers are able to generate X-ray haloes of elliptical merger remnants. In contrast to this investigations of cold gaseous discs, \citet{SiHo09} simulated the collisions of galaxies including hot gas within the galactic haloes. However, their models did not contain any disc gas. In their simulations of different major and minor merger (3:1 and 10:1 mass ratio) scenarios with different gas fractions and orbital properties, the gas in the haloes gets shock-heated already before the first encounter. The strongest shocks and highest X-ray emissions occur after the first encounter of the galaxies. They also find that the peak X-ray luminosity increases with increasing progenitor masses and also with increasing gas fraction.

Since the increase of the IGM temperature also depends on the initial magnetic field strength \citep{KoLe11}, a connection between the X-ray emission and the magnetization may be expected, but detailed investigations concerning this interconnection are still missing. Also, there are no studies of the X-ray emission during galactic interactions which include both gaseous discs and a sourrounding hot IGM gas.

Therefore, within the presented work we want to address the following issues: 
Until now, simulations of galactic mergers including magnetic fields have only been performed for major merger scenarios. However, as minor mergers are expected to be the more frequent events, we aim to study the effect of the mass ratio of the progenitor galaxies on the magnetic field evolution. Based on 32 different minor merger scenarios, we investigate the amplification of a given initial magnetic field within the galaxies and an ambient IGM numerically using the MHD version of \textsc{Gadget} \citep{DoSt09}. A particular focus is thereby placed on the evolution of the expected X-ray emission of the simulated systems. We also discuss the resolution-dependent numerical divergence in our simulations.

The paper is organized as follows: In section 2 we briefly describe the numerical method. The galaxy models and merger scenarios are described in detail in section 3. In section 4 we present the results of our simulated merger scenarios, particularly the magnetic field evolution and X-ray emission. Finally, we summerize and discuss our findings in section 5.

\section[]{Method}
All galaxy collision simulations have been performed with the N-body/SPH code \textsc{Gadget} \citep{SpYo01,Sp05}.
This code models dark matter and stars as a self-gravitating, collisionless fluid, which is treated with the traditional 
N-body approach. Gravitational interactions are computed with a Barnes \& Hut tree construction \citep{BaHu86}. 
The IGM and interstellar medium (ISM) gas is described as a conductive, ideal fluid, whereby
hydrodynamics are treated with the SPH method (for recent reviews see
e.g. \citealt{Ro09,Sp10,Pr11}). The applied SPH formulation conserves both energy and entropy \citep{SpHe02}. An additional MHD implementation by \citet{DoSt09} allows for the evolution of magnetic fields.
Below, we briefly describe the SPH and smoothed particle magnetohydrodynamics (SPMHD) methods. For detailed descriptions see e.g. \citet{DoSt09} and \citet{Pr11}.

\subsection{SPH}
In SPH, the fluid is discretized by mass elements $m_{i}$ at positions $\textbf{\textit{x}}_{i}$. The density can then be estimated using
\begin{equation}
\rho_{i} = \sum_{j=1}^{N} m_{j} W(\textbf{\textit{x}}_{i} - \textbf{\textit{x}}_{j},h_{i}),
\end{equation}
where the kernel $W$ \citep{MoLa85} is a weighting function determining the contribution of the particles inside a smoothing sphere with the radius $h_{i}$:
\begin{equation}
\begin{small}W(x,h) = \frac{8}{\pi h^{3}}
\left\{
\begin{aligned}
&1-6 \left( \frac{x}{h} \right)^{2} + 6 \left( \frac{x}{h}\right)^{3} && 0 \leq \frac{x}{h} < 0.5,\\
&2 \left( 1-\frac{x}{h}\right) ^{3}  && 0.5 \leq \frac{x}{h} < 1,\\
&0  &&1 \leq \frac{x}{h}.
\end{aligned}
\right.\end{small}
\end{equation}
To enable spatial adaptivity, the smoothing length $h_{i}$ is evaluated iteratively together with the density via
\begin{equation}
h(x_i)=\eta \left( \frac{m_{i}}{\rho_{i}}\right) ^{1/d},
\end{equation}
whereby the parameter $\eta$ is defined by the dimension $d$ of the problem and the shape of the kernel, within our 3-dimensional simulations by a cubic kernel:
\begin{equation}
N=\frac{4}{3}\pi \eta^{3},
\end{equation}
with the number of neighbouring particles $N$.

The equation of motion for the SPH particles can be derived using a variational principle, which leads to 
\begin{equation}
\left( \frac{\text{d}\textbf{\textit{v}}_{i}}{\text{d}t}\right)^{\text{hyd}} = - \sum_{j} m_{j} \left[  f_{i}^{\text{co}} \frac{P_{i}}{\rho_{i}^{2}} \nabla_{i} W_{i} + f_{j}^{\text{co}} \frac{P_{j}}{\rho_{j}^{2}} \nabla_{i} W_{j}\right],
\end{equation}
with  $W_{i}=W(\textbf{\textit{x}}_{i} -
\textbf{\textit{x}}_{j},h_{i})$ and  $W_{j}=W(\textbf{\textit{x}}_{i} -
\textbf{\textit{x}}_{j},h_{j})$. The correction terms $f_{i}^{\text{co}}$ account for the adaptive smoothing lengths:
\begin{equation}
f_{i}^{\text{co}}=\left[ 1+\frac{h_{i}}{3\rho_{i}}\frac{\partial \rho_{i}}{\partial h_{1}}\right] ^{-1}.
\end{equation}
We use the standard implementation of the artificial viscosity with values of $\alpha=2$ and $\beta=1.5$ for the dimensionless parameters (for details see \citealt{Pr11}).

\subsection{SPMHD}

In ideal MHD, the evolution of a magnetic field can be followed using the induction equation
\begin{equation}
\frac{\text{d}\textbf{\textit{B}}}{\text{d}t} =  \left( \textbf{\textit{B}} \cdot \textbf{$\nabla$}\right)  \textbf{\textit{v}} - \textbf{\textbf{\textit{B}}} \left( \textbf{$\nabla$} \cdot \textbf{\textit{v}}\right),
\end{equation}
whereby the constraint $\textbf{$\nabla$} \cdot \textbf{\textit{B}} = 0 $ has been used. The SPH discretization of the induction equation reads
\begin{equation}
\frac{\text{d}\textbf{\textit{B}}_{i}}{\text{d}t} = f_{i}^{\text{co}} \frac{1}{\rho_{i}} \sum_{j} m_{j} [ \textbf{\textit{B}}_{i} (\textbf{\textit{v}}_{ij} \cdot \nabla_{i}W_{i})  - \textbf{\textit{v}}_{ij} ( \textbf{\textit{B}}_{i} \cdot \nabla_{i}W_{i} )].
\end{equation}
We also apply artificial magnetic dissipation as described in \cite{DoSt09} which is based on \citet{PrMo04a}.

The magnetic field exerts a feedback on the plasma via the Lorentz force. The magnetic contribution to the acceleration of a particle can be written in the following symmetric and conservative form 
\begin{equation}
\left( \frac{\text{d}\textbf{\textit{v}}^{k}}{\text{d}t}\right) ^{\text{mag}} = \frac{1}{\rho} \frac{\partial M^{kl}}{\partial \textbf{\textit{x}}^{l}},
\end{equation}
with the magnetic stress tensor \citep{PhMo85}
\begin{equation}
M^{kl}_{i} := \frac{1}{\mu_{0}}\left( \textbf{\textit{B}}^{k}_{i} \textbf{\textit{B}}^{l}_{i} - \frac{1}{2} \textbf{\textit{B}}_{i}^{2}\delta^{kl}\right).
\end{equation}
Hence, the contribution to the equation of motion caused by magnetic fields can be discretized as follows \citep{PrMo04b}
\begin{equation}
\left(\frac{\text{d}\textbf{\textit{v}}_{i}^{k}}{\text{d}t}\right)^{\text{mag}} = \sum_{j}  m_{j} \left[f_{i}^{\text{co}} \frac{M_{i}^{kl}}{\rho_{i}^{2}} \nabla_{i}^{l} W_{i} + f_{j}^{\text{co}} \frac{M_{j}^{kl}}{\rho_{j}^{2}}\nabla_{j}^{l} W_{j}\right].
\end{equation}

However, for strong magnetic forces, this momentum conserving form can lead to
numerical instabilities, i.e. clumping of the particles \citep{PhMo85},
because of the non-vanishing numerical divergence of $\textbf{B}$, which will not be stabilized in cases where the magnetic pressure exceeds the gas pressure.
One method to circumvent this problem was suggested by \citet{BoOm01}. Within this method, the effects of any unphysical source terms of B are subtracted from the momentum equation. The
corresponding correction term can be calculated via
\hspace*{-5mm}
\begin{eqnarray}
\left( \frac{\text{d}\textbf{\textit{v}}_{i}}{\text{d}t}\right)^{\text{corr}}= \hspace*{55mm}\nonumber \\ \hspace*{-1mm}\frac{\textbf{\textit{B}}_{i}}{\mu_{0}} \hat{\beta} \sum_{j}  m_{j} \Bigg{[}   f_{i}^{\text{co}} \frac{\textbf{\textit{B}}_{i}}{\rho_{i}^{2}}  \nabla_{i}W_{i} + f_{j}^{\text{co}} \frac{\textbf{\textit{B}}_{j}}{\rho_{j}^{2}}  \nabla_{j}W_{j}\Bigg{]},
\end{eqnarray}
where $\hat{\beta}$ is a constant factor, with a typical value of
$\hat{\beta}=1$. Additionally, a threshold for the divergence force
subtraction was introduced in order to account for situations in which the acceleration
due to the divergence force could become dominant (\citealt{KoLe11}, Stasyszyn
\& Dolag 2011, in preparation). This threshold is set to half of the value of
the current Lorentz force. Generally, it was found that the effects due to the
violation of momentum conservation by the divergence force subtraction are of an insignificant order, whereas the implementation of the correction term significantly improves the results in test cases \citep{DoSt09}.

The MHD version of \textsc{Gadget} was already successfully employed for the analysis of magnetic fields in star formation, in isolated galaxies and mergers of galaxies as well as within galaxy clusters \citep[e.g.][]{BuCl11b,KoLe09,DoDo09,BoDo11}.

\subsection{Radiative cooling, star formation and supernova feedback}

Radiative cooling is modeled assuming the gas to be optically thin and in collisional ionization equilibrium \citep{KaWe96}. The gas is assumed to be of primordial composition (hydrogen fraction X=0.76, helium fraction Y=0.24), hence, metal-dependent effects are neglected. This approach is reasonable as we will consider the primary mechanism of X-ray emission (see chapter 4.3) to be thermal bremsstrahlung, which is consistent with the application of zero-metallicity cooling \citep{CoDi06}. Also, we account for a homogeneous extragalactic ultraviolet (UV) background in our simulations (e.g. \citealt{HaMa96}).

Star formation (SF) and associated supernova feedback are treated by the hybrid multiphase model as described in \citet{SpHe03}, where the interstellar medium is modeled as a two-phase fluid \citep{McOs77}. This fluid is assumed to consist of cold clouds which are
embedded in an ambient hot medium. At high densities, material gets bound in
these clouds and star formation can take place, which in turn lowers
the density of the ambient gas phase. This leads to a decrease of its radiative
losses and enables the heating of the region caused by supernova explosions, resulting in pressure support for the ISM. The multiphase model is realized numerically by describing each star-forming particle to be made up of two components, a
hot diffuse component and a cold component, whereby the local gas density and temperature determine the relative amounts of the
phases. The effective equation of state (EOS) for the gas is given by $P_{\text{eff}} = (\gamma -1)(\rho_{\text{h}}u_{\text{h}} + \rho_{\text{c}}u_{\text{c}})$ \citep{SpHe03,SpDi05b}, with the density and the specific thermal energy of the hot ($\rho_{\text{h}}$ and $u_{\text{h}}$) and the cold phase ($\rho_{\text{c}}$ and $u_{\text{c}}$), and an adiabatic index of $\gamma = 5/3$.

\section[]{Initial Conditions}
\subsection{Galaxy models}
In order to perform a series of simulations of unequal mass mergers, we set up structurally similar galaxy models with different total masses. The galaxies are set up using the method described by \citet{SpDi05b}. This method allows for a galaxy model consisting of a cold dark matter halo, an exponential stellar disc, a stellar bulge (all of these components being collisionless N-body particles) and an exponential gaseous disc (SPH particles).
Both the virial mass $M_{200}$ and the virial radius $r_{200}$ of the galaxy depend on the virial velocity $v_{200}$ \citep{SpDi05b,JoNa09} via
\begin{equation}
r_{200} = \frac{v_{200}}{10 H_{0}},
\end{equation}
\begin{equation}
M_{200} = \frac{v_{200}^{3}}{10 G H_{0}},
\end{equation}
with the Hubble constant H$_{0}$=$h \cdot 100$ km s$^{-1}$ Mpc$^{-1}$ and $h=0.71$. Thus, we determine the masses of our
galaxy models by varying the virial velocity $v_{200}$. We assume the same
halo concentration CC, spin parameter $\lambda$, disc and bulge mass fractions
$m_{\text{d}},m_{\text{b}}$, disc spin fraction $j_{\text{d}}$, gas fraction
f, disc height $z_{0}$, bulge size $l_{\text{b}}$ and the scale length of extended
gas disc $l_{\text{g}}$ for all of the used galaxy models. These common parameters and the parameters of the multi-phase model are given in Table 1, where masses are given in units of the total galactic mass M$_{\text{tot}}$ and scale lengths in units of the stellar disc scale length $l_{\text{d}}$.

\begin{table}
\caption{Parameters of initial setup common to all galaxy models}
\begin{center}
\renewcommand{\arraystretch}{1.2}
\begin{tabular}{lll}
\hline\hline  
\multicolumn{3}{c}{\textsc{Disc Parameters}} \\ \hline\hline
Concentration&CC&12\\
Spin parameter&$\lambda$&0.1\\
Disk mass fraction$^{a}$&$m_{\text{d}}$&0.05 $\cdot$ M$_{\text{tot}}$\\ 
Bulge mass fraction$^{a}$&$m_{\text{b}}$&0.02 $\cdot$ M$_{\text{tot}}$\\
Disk spin fraction&$j_{\text{d}}$&0.05\\
Gas fraction&f&0.2\\
Disk height$^{a}$&$z_{0}$&0.2 $\cdot$ $l_{\text{d}}$\\
Bulge size$^{a}$&$l_{\text{b}}$&0.2 $\cdot$ $l_{\text{d}}$\\
Scale length of extended gas disc$^{a}$&$l_{\text{g}}$&6 $\cdot$ $l_{\text{d}}$\\\hline\hline

&\hspace*{-2.8cm}\textsc{Multi Phase Model Parameters}&\\\hline\hline
Gas consumption timescale&t$_{\text{MP}}$&8.4 Gyr\\
Mass fraction of massive stars&$\beta_{\text{MP}}$&0.1\\
Evaporation parameter&A$_{0}$&4000\\
Effective supernova temperature&T$_{\text{SN}}$&4 $\cdot$ $10^{8}$ K\\
Temperature of cold clouds&T$_{\text{CC}}$&1000 K\\\hline
\multicolumn{3}{l}{\scriptsize{(a) M$_{\text{tot}}$ and l$_{\text{d}}$ are given in Table 3.}}
\end{tabular}
\end{center}
\label{1}
\end{table}

\begin{table*}
 \begin{minipage}{126mm}
  \caption{Galaxy Model Parameters}
\begin{center}
\renewcommand{\arraystretch}{1.2}
\begin{tabular}{l|rrr|rrrr}
\hline\hline  
\multicolumn{8}{c}{\textsc{medium resolution}} \\ \hline\hline
Model&M$_{\text{tot}}$&R$_{200}$&\hspace{0.15cm}$l_{\text{d}}$&$N_{\text{halo}}^{\text{a}}$ & $N_{\text{disc}}^{\text{b}}$ & $N_{\text{gas}}^{\text{c}}$ & $N_{\text{bulge}}^{\text{d}}$\\
& \begin{footnotesize}[$10^{10}M_{\sun}$] \end{footnotesize}&\begin{footnotesize}[kpc/$h$]\end{footnotesize} &\begin{footnotesize}[kpc/$h$]\end{footnotesize} &&&&\\\hline

M1&134.14&160.0&7.09&400 000&960 000&240 000&400 000\\
M2&63.96&125.0&5.54&200 000&457 764&114 441&190 735\\
M3&43.60&110.0&4.87&133 333&311 954&77 988&129 981\\
M4&32.75&100.0&4.24&100 000&234 375&58 594&97 656\\
M5&26.77&93.5&4.14&80 000&191 579&47 895&79 824\\
M6&22.32&88.0&3.90&66 667&159 720&39 930&66 550\\
M7&19.07&83.5&3.70&57 143&136 449&34 112&56 854\\
M8&16.77&80.0&3.54&50 000&120 000&30 000&50 000\\
M9&14.95&77.0&3.41&44 444&107 000&26 750&44 583\\
M10&13.27&74.0&3.28&40 000&94 974&23 744&39 573\\
M15&8.99&65.0&2.88&26 667&64 365&16 091&26 819\\
M20&6.73&59.0&2.61&20 000&48 136&12 034&20 057\\\hline\hline

\multicolumn{8}{c}{\textsc{high resolution}} \\ \hline\hline

M1&134.14&160.0&7.09&4 000 000&9 600 000&2 400 000&4 000 000\\
M3&43.60&110.0&4.87&1 333 333&3 119 536&779 884&1 299 807\\
M50&1.91&43.5&1.93&80 000&192 921&48 230&80 384\\
M100&0.96&34.5&1.53&40 000&96 243&24 061&40 101\\\hline\hline  

\multicolumn{8}{c}{\textsc{low resolution}} \\ \hline\hline

M1&134.14&160&7.09&40 000&96 000&24 000&40 000\\
M3&43.60&110&4.87&13 333&31 195&7 799&12 998\\\hline

\multicolumn{8}{l}{\scriptsize{(a) collisionless particles within dark matter halo \hspace{0.3cm} (b) collisionless particles within disc}}\\ \multicolumn{8}{l}{\scriptsize{(c) gas particles within disc \hspace{2.75cm}  (d) collisionless particles within bulge}}\\
\end{tabular}
\end{center}
\end{minipage}
\label{3}
\end{table*}

We set up 12 medium resolution galaxy models, 4 high resolution models and two
low resolution models. Model M1 is the most massive galaxy in our sample, and
the other models are numbered according to their mass with respect to the M1
model, i.e. model M2 is a galaxy with half the mass of model M1, M3 has one
third of the mass of M1 and so forth\footnote{By varying the virial velocity it is not possible to set the mass of the galaxies to an exact predefined value. Therefore, our smaller models contain only approximately a certain percentage of the model M1, which is reflected in the irregular particle numbers in Table 2.}. This allows for a transparent merger
denotation, i.e. merger M1M2 describes a 2:1 merger, M1M3 a 3:1 merger and so
on. We refer to the mass ratios in terms of the mass of the companion galaxy divided by the mass of M1. 

The used particle numbers, total masses, virial radii and stellar disc scale lengths used for our different galaxy models are summarized in
Table 2. These setups result in a mass of the gas particles of  $m_{\text{gas}}\approx4\cdot10^{4}$ $h^{-1}$ M$_{\sun}$
for the medium resolution,
$m_{\text{gas}}\approx4\cdot10^{3}$ $h^{-1}$ M$_{\sun}$ for
the high resolution and $m_{\text{gas}}\approx4\cdot10^{5}$
$h^{-1}$ M$_{\sun}$ for the low resolution. For the medium resolution, the fixed gravitational softening
lengths are $\epsilon = 0.11$ $h^{-1}$kpc for the dark matter particles and
$\epsilon = 0.022$ $h^{-1}$kpc for the gas, bulge and halo particles. For the low and high resolution runs, these softening lengths have been adjusted using $h_{\text{new}} = h_{\text{old}} \cdot (N_{\text{old}}/N_{\text{new}})^{1/3}$ (see e.g. \citealt{De01,JoNa09}). The gravitational softening lengths are summarized in Table 3. The minimum SPH smoothing length for the gas particles is 1.0$\epsilon$.

\begin{table}
\caption{Gravitational softening lengths $\epsilon$}
\begin{center}
\renewcommand{\arraystretch}{1.2}
\begin{tabular}{lllll}
\\ \hline\hline
&Halo&Disk&Gas&Bulge\\
&[pc/h]&[pc/h]&[pc/h]&[pc/h]\\\hline
low resolution&240&48&48&48\\
medium resolution&110&22&22&22\\
high resolution&52&10&10&10\\
\hline
\end{tabular}
\end{center}
\end{table}

\begin{table*}
\caption{Simulated galaxy merger scenarios}
\begin{center}
\renewcommand{\arraystretch}{1.2}
\begin{tabular}{llllll}
\hline\hline
\multicolumn{6}{c}{\textsc{Merger scenarios}} \\ \hline\hline
Scenario&mass ratio&orbit&initial $B_{\text{gal},0}$&initial $B_{\text{IGM},0}$&resolution\\\hline
M1M2\_G9I9&2:1&prograde&10$^{-9}$ G&10$^{-9}$ G&medium\\
M1M3\_G9I9&3:1&prograde&10$^{-9}$ G&10$^{-9}$ G&medium\\
M1M4\_G9I9&4:1&prograde&10$^{-9}$ G&10$^{-9}$ G&medium\\
M1M5\_G9I9&5:1&prograde&10$^{-9}$ G&10$^{-9}$ G&medium\\
M1M6\_G9I9&6:1&prograde&10$^{-9}$ G&10$^{-9}$ G&medium\\
M1M7\_G9I9&7:1&prograde&10$^{-9}$ G&10$^{-9}$ G&medium\\
M1M8\_G9I9&8:1&prograde&10$^{-9}$ G&10$^{-9}$ G&medium\\
M1M9\_G9I9&9:1&prograde&10$^{-9}$ G&10$^{-9}$ G&medium\\
M1M10\_G9I9&10:1&prograde&10$^{-9}$ G&10$^{-9}$ G&medium\\
M1M15\_G9I9&15:1&prograde&10$^{-9}$ G&10$^{-9}$ G&medium\\
M1M20\_G9I9&20:1&prograde&10$^{-9}$ G&10$^{-9}$ G&medium\\\hline
M1M2\_G9I12&2:1&prograde&10$^{-9}$ G&10$^{-12}$ G&medium\\
M1M3\_G9I12&3:1&prograde&10$^{-9}$ G&10$^{-12}$ G&medium\\
M1M4\_G9I12&4:1&prograde&10$^{-9}$ G&10$^{-12}$ G&medium\\
M1M5\_G9I12&5:1&prograde&10$^{-9}$ G&10$^{-12}$ G&medium\\
M1M6\_G9I12&6:1&prograde&10$^{-9}$ G&10$^{-12}$ G&medium\\
M1M7\_G9I12&7:1&prograde&10$^{-9}$ G&10$^{-12}$ G&medium\\
M1M8\_G9I12&8:1&prograde&10$^{-9}$ G&10$^{-12}$ G&medium\\
M1M9\_G9I12&9:1&prograde&10$^{-9}$ G&10$^{-12}$ G&medium\\
M1M10\_G9I12&10:1&prograde&10$^{-9}$ G&10$^{-12}$ G&medium\\
M1M15\_G9I12&15:1&prograde&10$^{-9}$ G&10$^{-12}$ G&medium\\
M1M20\_G9I12&20:1&prograde&10$^{-9}$ G&10$^{-12}$ G&medium\\\hline
M1M50\_G9I9-hr&50:1&prograde&10$^{-9}$ G&10$^{-9}$ G&high\\
M1M100\_G9I9-hr&100:1&prograde&10$^{-9}$ G&10$^{-9}$ G&high\\\hline
M1M4\_G6I9&4:1&prograde&10$^{-6}$ G&10$^{-9}$ G&medium\\
M1M7\_G6I9&7:1&prograde&10$^{-6}$ G&10$^{-9}$ G&medium\\\hline
M1M3\_G9I9-ret&3:1&retrograde&10$^{-9}$ G&10$^{-9}$ G&medium\\
M1M7\_G9I9-ret&7:1&retrograde&10$^{-9}$ G&10$^{-9}$ G&medium\\\hline
M1M3\_G9I9-lr&3:1&prograde&10$^{-9}$ G&10$^{-9}$ G&low\\
M1M3\_G9I9-hr&3:1&prograde&10$^{-9}$ G&10$^{-9}$ G&high\\\hline
M1M4\_G0I0&4:1&prograde&0 G&0 G&medium\\
M1M7\_G0I0&7:1&prograde&0 G&0 G&medium\\\hline
\hline
\end{tabular}
\end{center}
\label{2}
\end{table*}

\subsection{Galactic magnetic field}
The initial magnetization of the progenitor discs is given by
$B_{\text{x}}=B_{\text{gal,0}}$ and $B_{\text{y}}=B_{\text{z}}=0$ G with the
z-axis being the axis of rotation. Thus, the initial field
lies always in the plane of the galactic discs. As we are interested in the influence of minor mergers on the galactic magnetic field evolution, particularly its amplification, we mainly focus on a small initial
magnetic field strength of $B_{\text{gal,0}} = 10^{-9}$ G. This value is by three orders of magnitude smaller than the typical observed galactic magnetic field value \citep{BeBr96}. However, we also simulate two of our merger scenarios (see section 3.6.) with a initial magnetic field strength of $B_{\text{gal,0}} = 10^{-6}$ G as well as mergers without any magnetic fields for comparison.

\subsection{Orbits}

Within all of the presented simulations the largest (i.e. most massive) galaxy model M1 interacts with one of the smaller (i.e. less massive) models M2-M100. As only minor mergers are considered, we will refer to the smaller galaxy in each simulation as the \textquotedblleft companion galaxy\textquotedblright. 
In order to ensure a collision, the galaxies are set on a parabolic orbit, resulting in a prograde encounter (i.e. the spin direction is the same within both galaxies) in most of the simulations. Two of the  models are set on a retrograde encounter for comparison. The initial separation $r_{\text{sep}}$
of the galaxies is determined by the sum of their virial radii. The pericenter distance in all simulations is $r_{\text{p}} = 5$ $h^{-1}$kpc. The disc orientation (see \citealt{ToTo72}) is set to $\iota=60^{\circ}$ and $\omega=60^{\circ}$ for the largest galaxy M1 and $\iota=60^{\circ}$ and $\omega=-60^{\circ}$ for the companion galaxy, respectively, within all of our merger scenarios. A detailed description of these merger scenarios is given in section 3.6.

\begin{figure*}
 \begin{minipage}{\textwidth}
	\begin{center}
\includegraphics[width=\textwidth]{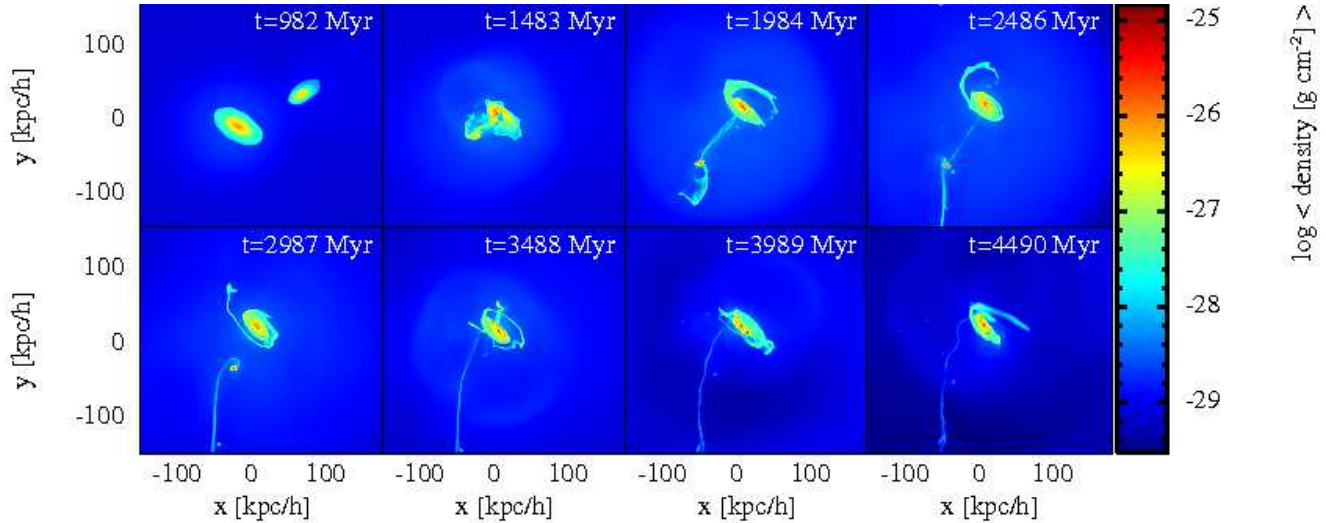}
	\end{center}
\vspace*{-6.4cm}
\end{minipage}
\caption{\small{ Evolution of the projection of the mean line-of-sight density at eight different time steps (given within the panels) for the minor merger scenario M1M4$\_$G6I9.}}
\label{dens1}
\end{figure*}

\subsection{IGM}
We include an ambient IGM composed of additional gas particles surrounding the
galaxies similar to \citet{KoLe11}. The IGM particles are arranged in a hexagonal
closed-packed lattice. The mass of the IGM gas particles is the same as the mass of the galactic gas particles. The volume filled with the IGM is 700 \texttimes 700 \texttimes 700 $h^{-3}$ kpc$^{3}$ centered on the common center of mass of the progenitor galaxies at the beginning of the simulation. For simplicity, we assume that the IGM is pervading the galaxies.
The density of the IGM is $\rho_{\text{IGM}}=10^{-29}$ g cm$^{-3}$, resulting in IGM particle numbers
of $N _{\text{IGM}}$= 89 713 for the low, $N_{\text{IGM}}$= 905 205 for the medium and $N_{\text{IGM}}= 8\hspace{1mm} 997 \hspace{1mm}083$ for the high resolution setups.

The internal energy of the IGM is calculated via $u_{\text{IGM}} = v_{200}^{2}/2$, using the virial velocity $v_{200}$ of the larger progenitor galaxy M1. This sets the temperature of the IGM to the virial temperature at the virial radius of the galaxy M1:
\begin{equation}
T_{\text{IGM}}=\frac{2}{3}u_{\text{IGM}}\frac{m_{p} \mu}{k_{B}} = \frac{1}{3}\langle v_{200}^{2}\rangle \frac{m_{p} \mu}{k_{B}} \approx 6 \cdot 10^{5} \hspace{1mm} \text{K},
\end{equation}
with the mean molecular weight for a fully ionized gas of primordial
composition $\mu \approx 0.588$, proton mass $m_{p}$ and Boltzmann constant $k_{B}$.

\subsection{IGM magnetic field}
The initial magnetic field of the IGM is assumed to be homogeneous and directed in x-direction, i.e. $ B_{\text{IGM,0}} = B_{\text{IGM,x}}$, with now the x-y-plane being the orbital plane. Due to our setup, the IGM magnetic field is also pervading the galaxies. $B_{\text{IGM,0}}$ is assumed to have values of either $10^{-9}$ G or
$10^{-12}$ G, respectively, depending on the merger scenario (see the next section). For the runs excluding magnetic fields also the IGM is not magnetized.

\subsection{Merger scenarios}
We have performed 11 standard, medium-resolution minor merger simulations with a standard galactic and IGM magnetic field of $B_{\text{gal,0}} = B_{\text{IGM,0}} = 10^{-9}$ G and a prograde disc orientation. The mass ratios of the discs range from 2:1 down to 20:1. The denotation \textquotedblleft G9I9\textquotedblright \hspace{1mm} in our scenario titles refers to the exponent of the initial galactic (G) and IGM (I) magnetic field strengths. We resimulate all of these 11 merger scenarios with a lower IGM magnetic field of $B_{\text{IGM,0}}=10^{-12}$ G. We also perform two additional simulations with  mass ratios of 50:1 and 100:1, respectively. For these two simulation we have to adopt a high resolution in order to sample the companion galaxy with at least 10 000 gas particles. Two of the standard scenarios (mass ratios 4:1 and 7:1) are additionally resimulated with a higher galactic magnetic field of $B_{\text{gal,0}}=10^{-6}$ G (we will refer to these models also as to the present-day mergers). Also, we resimulate two of the standard scenarios (mass ratios 3:1 and 7:1) with a retrograde disc orientation. Additionally, we resimulate the 3:1 standard scenario with both low and high resolution for resolution studies. Finally, two of the standard scenarios (mass ratios 4:1 and 7:1) are resimulated excluding magnetic fields completely. Thus, we present altogether 32 merger scenarios in this paper. The different scenarios are listed in Table 4.

\section[]{Simulations}

We let our initial system evolve for 200 Myr to allow possible
numerical discontinuities associated with the initial setup (e.g. effects
caused by overlayed magnetic fields of galaxies and IGM) to relax.

In the following, the physical parameters shown in the plots are calculated by taking the mean value $\langle L \rangle = \frac{1}{N}\sum_{j}{L_{j}}$ of all particles, if not stated otherwise.

\begin{figure}
\vspace*{-5.7cm}
\hspace*{-3.2cm}
 \begin{minipage}{\textwidth}
	\begin{center}
\includegraphics[width=0.8\textwidth]{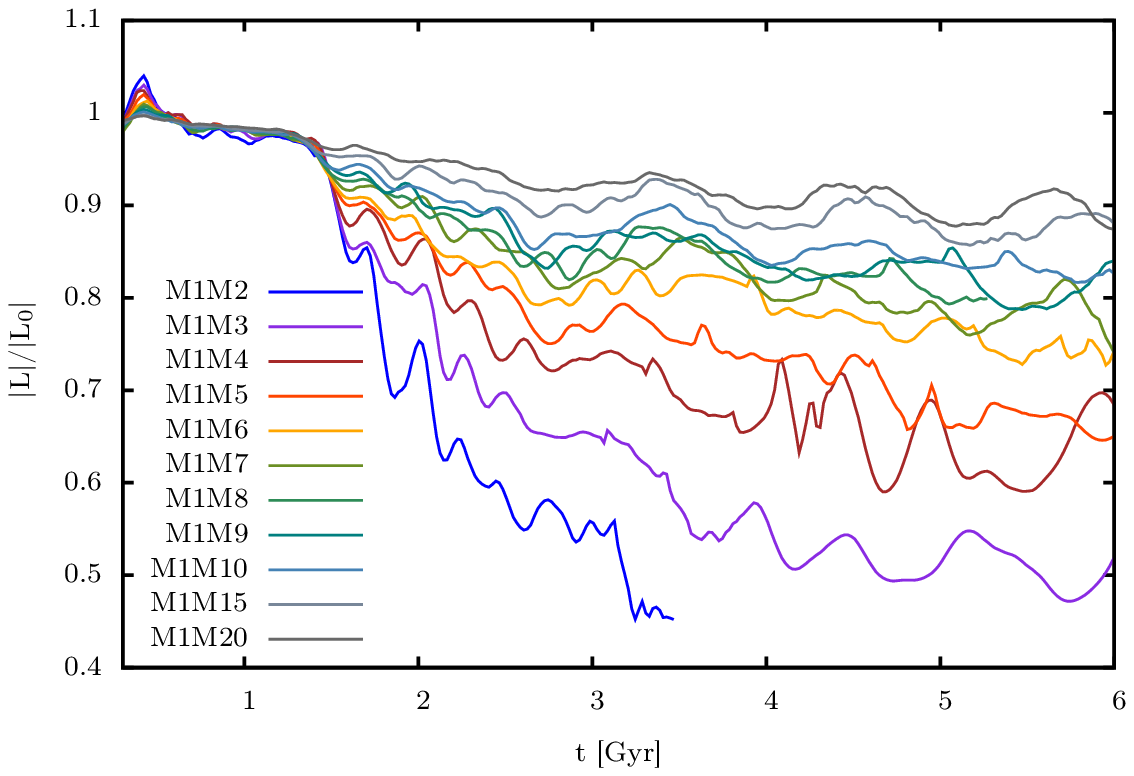}
	\end{center}
\end{minipage}
\vspace*{-9.5cm}
\caption{\small{ Angular momentum of the larger progenitor galaxy M1 as a function of time for the standard merger scenarios with $B_{\text{gal,0}} = B_{\text{IGM,0}} = 10^{-9}$ G. }}
\end{figure}

\subsection{Global evolution}

Fig. 1 shows the projection of the mean line-of-sight density exemplarily for the present-day scenario M1M4\_G6I9, whereby the 8 panels display a sequence of increasing time. 
In the beginning of the simulation, the galaxies are moving towards each other due to their mutual gravitational attraction. The first encounter takes place at $t\approx 1.3$ Gyr \footnote{All of the scenarios show the first encounter at $t\approx 1.3$ Gyr, but the first encounter for smaller mass ratios is slightly earlier than for collisions with larger mass ratios. The reason for this is the separation distance of the galaxies, which we set to the sum of the virial radii of the two galaxies. As the dark matter haloes already overlap at this distance, the Keplerian orbit is only nearly-parabolic.}, whereupon prominent tidal arms are developing. The first encounter is followed by a series of subsequent encounters until the final merger occurs at about $t\approx 4$ Gyr. For smaller progenitor galaxies, subsequent encounters take place at later times (and more of them until the final merger) due to the weaker gravitational attraction (not shown). For example, the final merger takes place at 3.4, 6.5, and 8.5 Gyr for the M1M2, the M1M7 and the M1M10 scenario, respectively. In most of the simulated scenarios, the disc of the larger progenitor galaxy M1 outlasts the interaction. However, in scenarios with the largest companion galaxies M2 and M3 (scenarios M1M2 and M1M3), the disc of M1 gets largely disrupted.

In order to quantify the amount of disc disruption, we estimate the angular momentum of the galaxy M1 by calculating $\textbf{L}= m (\textbf{r}\times \textbf{v})$ for all particles initially belonging to M1, with the \textbf{L}-axis being perpendicular to the galactic plane. Given this simple approach we may miss particles which are accreted onto the galaxy during the simulations, and spuriously account for particles which  have already left the disk. Hence, the calculated angular momentum cannot represent the true angular momentum perfectly. However, its evolution can nevertheless reveal the dynamical evolution of the disc.
Fig. 2 shows the normalized angular momentum $|\textbf{L}|/|\textbf{L}_{0}|$ (with $\textbf{L}_{0}$ the angular momentum at the beginning of the simulations) of M1 with time. At the time of the first encounter ($\approx 1.3$ Gyr), the angular momentum generally decreases, whereby the decrease is stronger for higher mass ratios. This is reasonable as larger companion galaxies can disrupt the disc of galaxy M1 more heavily. For the mergers with the largest mass ratios, only a slight decrease in the angular momentum is visible. 
The fluctuations of the angular momentum within all scenarios are due to the asymmetric mass distribution within the gaseous discs after the first encounter. They are not caused by further encounters. These fluctuations are more pronounced for larger companion galaxies, as the encounters have a stronger impact on the mass distribution.

\subsection{Magnetic field evolution}

The magnetic field is assumed to get enhanced through turbulent motions  \citep[see e.g.][for a review]{BrSu05}, which are caused by the impact and rotational energies of the galaxies. Therefore, we expect galaxy collisions with higher impact energies (i.e. galaxies with larger companion galaxies, see chapter 4.3.1) to be able to amplify the magnetic field more efficiently. Also, collisions of systems with different spin directions may be expected to lead to a more efficient amplification, as different rotational velocities can cause higher shear motions and could thus lead to a higher turbulence within the system.

Fig. 3 shows the morphological evolution of the projection of the mean line-of-sight total magnetic field again for the present-day scenario M1M4\_G6I9, whereby the 8 panels display a sequence of increasing time. Before the first encounter at about 1.3 Gyr, the galactic magnetic field gets
wound up by the differential rotation of the discs and gains a non-axisymmetric pattern with two magnetic arms (Fig. 3, left upper panel). Shocks and interaction-driven outflows caused by the first encounter are propagating into the IGM, whereby the IGM magnetic field is strengthened within the shocked regions (Fig. 3, second upper panel), an effect which has been also found by \citet{KoLe11}. During the subsequent evolution, further encounters take place, which are also accompanied by shocks and outflows and therefore lead to a further magnetization of the IGM. At the time of the final merger ($\approx$ 4 Gyr), the magnetic field in the galaxies (i.e. within the region with a density higher than $10^{-29}$ g cm$^{-3}$) has approximately retained its initial value of $10^{-6}$ G, whereas the IGM magnetic field got amplified within an extended region around the galaxies up to a value of several $10^{-9}$ G.

\begin{figure*}
\vspace*{-0.0cm}
 \begin{minipage}{\textwidth}
	\begin{center}
\includegraphics[width=1.\textwidth]{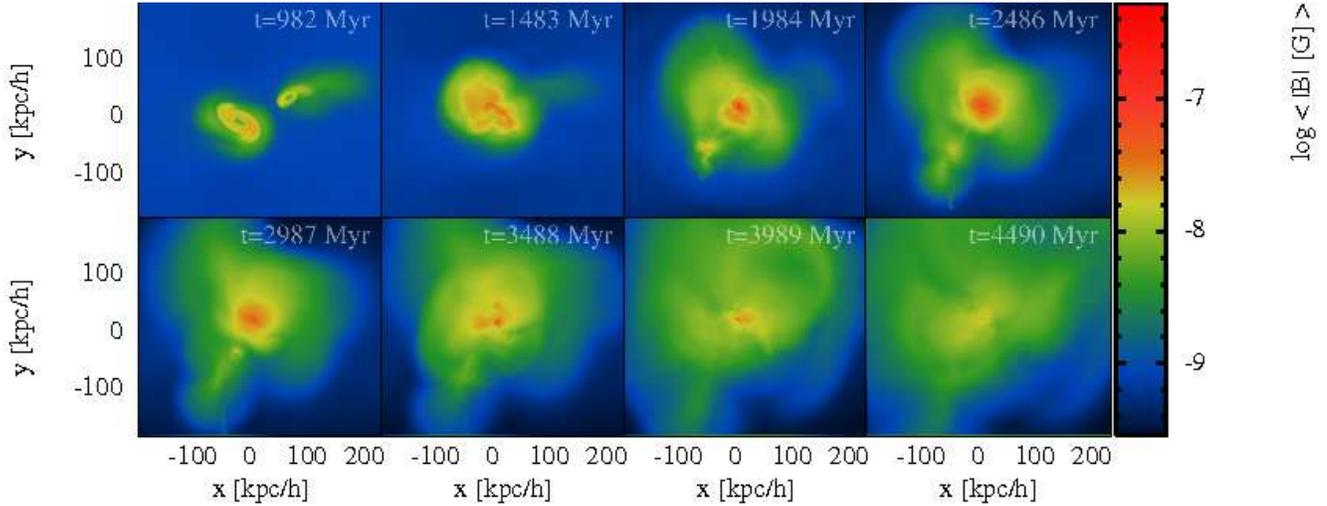}
	\end{center}
\vspace*{-6.5cm}
\caption{\small{ Evolution of the projection of the mean line-of-sight total magnetic field |\textbf{B}| at eight different time steps (given within the panels) for the minor merger scenario M1M4$\_$G6I9. }}
\end{minipage}
\end{figure*}

\begin{figure*}
 \begin{minipage}{\textwidth}
	\begin{center}
\includegraphics[width=\textwidth]{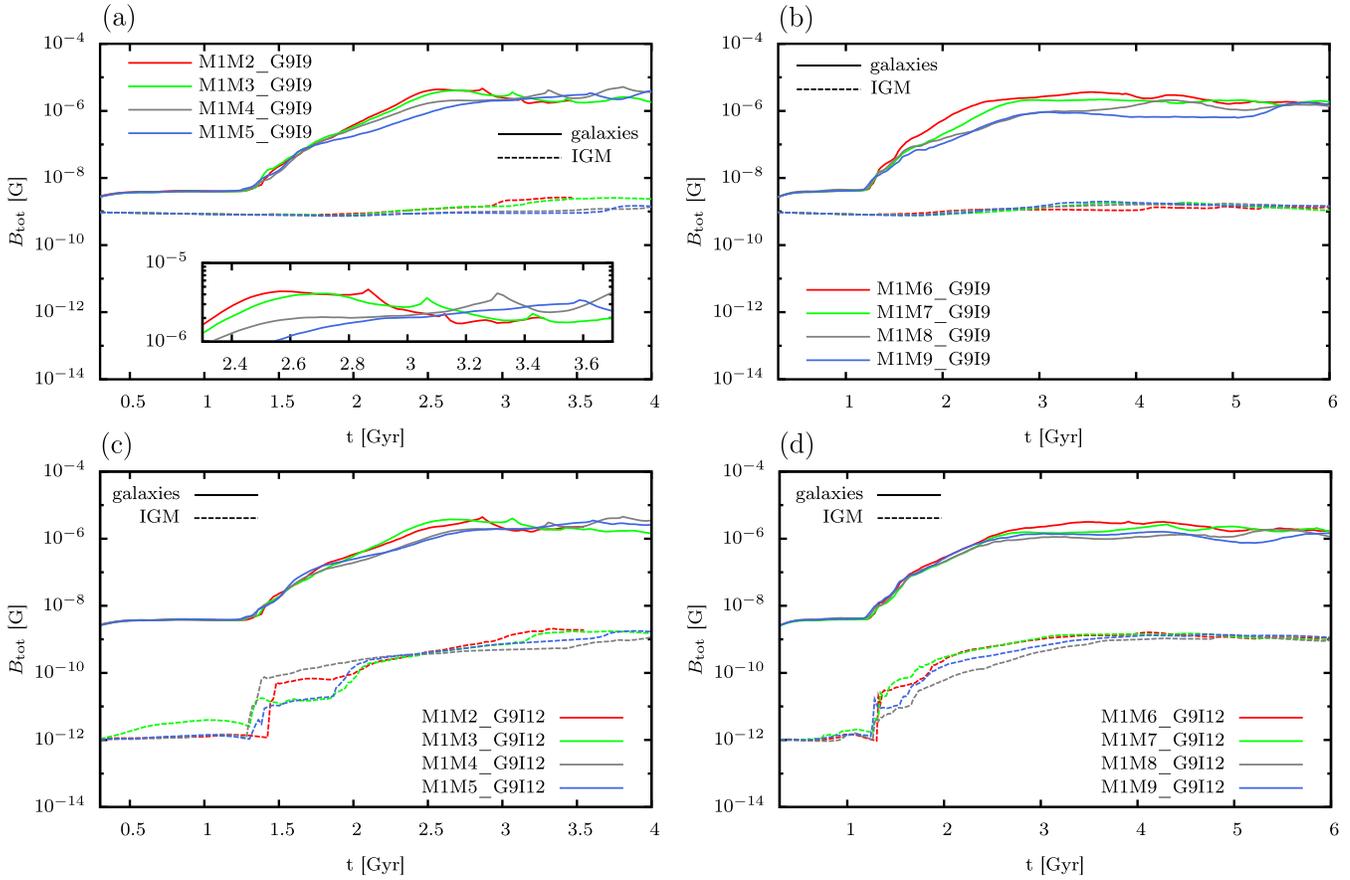}
	\end{center}
\caption{\small{ Total magnetic field $B_{\text{tot}} = \sqrt{B_{x}^{2} + B_{y}^{2} + B_{z}^{2}}$ as a function of time for the eight merger scenarios with the standard magnetization of $B_{\text{gal},0}=B_{\text{IGM},0}=10^{-9}$ G (\textit{upper panels}) and for the same eight scenarios but with a smaller initial IGM magnetic field of $B_{\text{IGM},0}=10^{-12}$ (\textit{lower panels}). Galaxies (solid lines) and IGM (dashed lines) are shown separately using a density threshold of $10^{-29}$ g cm$^{-3}$. The evolution of the magnetic field is similar for all merger scenarios, independent of the mass ratios of the progenitor discs. However, for smaller companion galaxies, the maximum value of the total magnetic field is slightly lower, and the slope of the amplification is slightly flatter compared to the scenarios with larger companion galaxies. }}
\end{minipage}
\end{figure*}

\begin{figure*}

 \begin{minipage}{\textwidth}
	\begin{center}
\includegraphics[width=\textwidth]{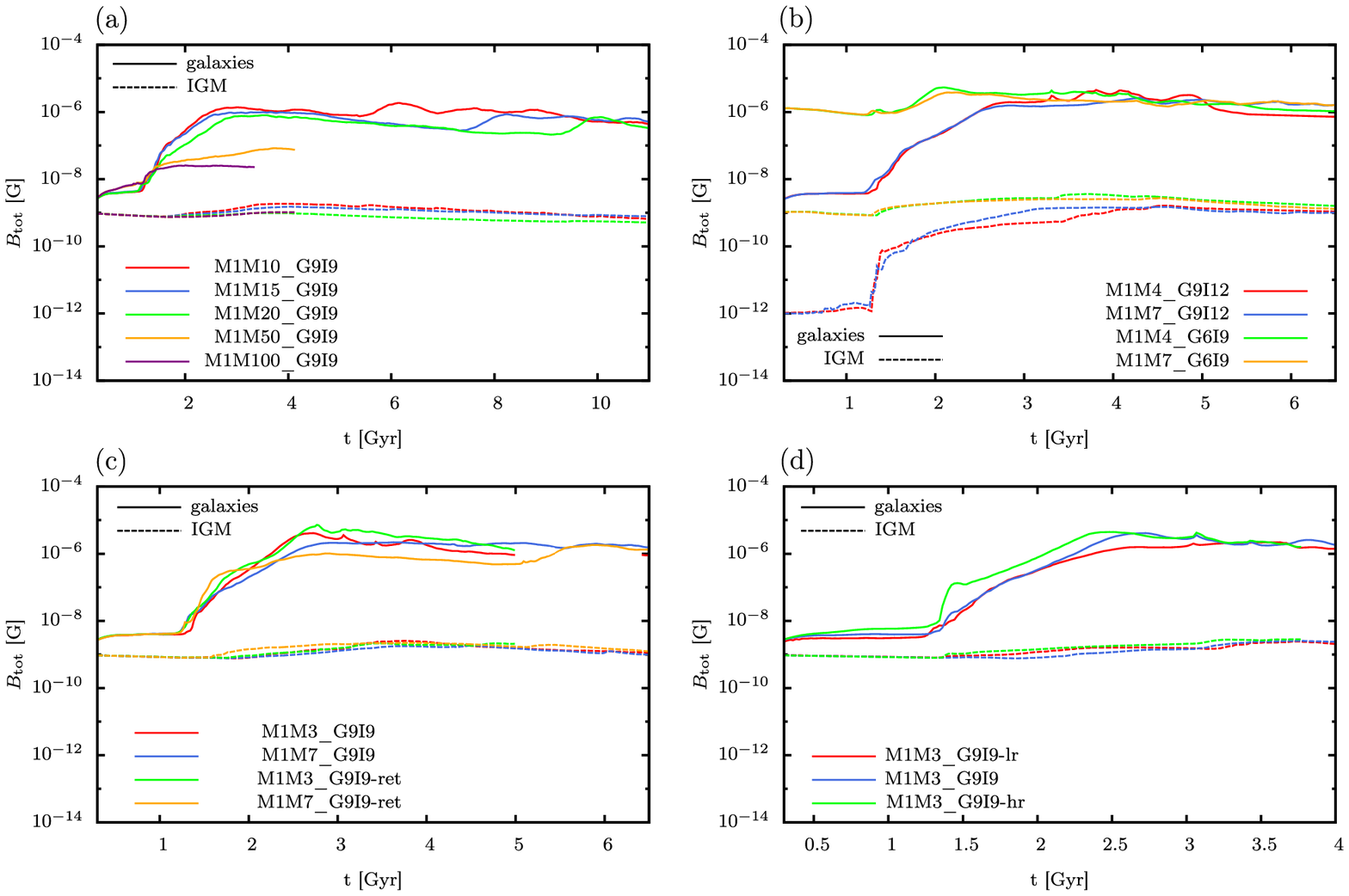}
	\end{center}

\end{minipage}
\caption{\small{ Evolution of the total magnetic field $B_{\text{tot}} = \sqrt{B_{x}^{2} + B_{y}^{2} + B_{z}^{2}}$ as a function of time for different merger scenarios with different initial magnetic fields, disc orientations and resolutions. (a) This panel shows the magnetic field evolution for the five merger scenarios with the smallest companion galaxies and with the standard magnetization of $10^{-9}$ G. (b) Magnetic field evolution of the merger scenarios M1M4 and M1M7, with the highest magnetization of $B_{\text{gal,0}} = 10^{-6}$ G, $B_{\text{IGM,0}}=10^{-9}$ G and also with the smallest magnetic field of $B_{\text{gal,0}} = 10^{-9}$ and $B_{\text{IGM,0}}=10^{-12}$ G. (c) Magnetic field evolution as a function of time for the two retrograde scenarios M1M3 and M1M7 with the standard initial magnetization of $10^{-9}$ G together with the corresponding prograde scenarios. (d) Magnetic field evolution as a function of time for the scenario M1M3 for the three different resolutions. Galaxies (solid lines) and IGM (dashed lines) are again plotted separately using a density threshold of $10^{-29}$ g cm$^{-3}$.}}
\end{figure*}

\subsubsection{Dependence on the mass ratio}

Fig. 4 shows the evolution of the total magnetic field $B_{\text{tot}} = \sqrt{B_{x}^{2} + B_{y}^{2} + B_{z}^{2}}$ as a function of time for eight of the standard scenarios ($B_{\text{gal},0} = B_{\text{IGM},0} = 10^{-9}$ G, Fig. 4a \& b) and for the same scenarios with a lower initial IGM magnetic field of $B_{\text{IGM,0}} = 10^{-12}$ G (Fig. 4c \& d). We plot the galactic magnetic field (solid lines) and the IGM magnetic field (dotted lines) separately using a threshold of $10^{-29}$g cm$^{-3}$. 

In all of the presented simulations, a slight amplification of the total galactic magnetic field by approximately a factor of 4 caused by the winding process is visible in the two progenitor discs before the first encounter. This behaviour was already observed in simulations of isolated galaxies and galactic major mergers (\citealt{KoLe09,KoKa10,KoLe11}). All of the presented merger simulations show this initial amplification with the same order of magnitude.

During the first encounter, the galactic magnetic field gets efficiently amplified within all of the presented simulations, whereby the maximum magnetic field strength and the slope of the amplification depend on the mass ratio of the progenitor galaxies. Thereby, lower mass ratios lead to a slightly higher maximum value of the magnetic field and a steeper slope of the amplification, caused by the presumably higher turbulence driven by the first encounter. In case of lower mass ratios, the maximum value of the galactic magnetic field strength is mostly reached after the first encounter. However, the maximum value of the magnetic field in scenarios with higher mass ratios is usually not reached until the time of the second encounter.

The spikes in the galactic magnetic field strength after the maximum value in the simulations with mass ratios of 2:1 up to 5:1 (Fig. 4a (inset), 4c) correspond to the second encounters. The second encounter takes place at a later time for smaller companion galaxies (cf. section 4.1.). For smaller companion galaxies, the second encounter generally results in a further increase of the galactic magnetic field strength, whereby the maximum value of the magnetic field is reached at the time of the second encounter. However, in most of the merger scenarios, the galactic magnetic field decreases again after the second encounter. The subsequent encounters and the final merger do not lead to any considerable further amplification of the galactic magnetic field. At the end of the simulations, the galactic magnetic field strengths for the presented merger scenarios saturate at a similar value of several $\mu$G. The evolution of the IGM magnetic field is discussed separately in section 4.2.5.

Fig. 5 shows the evolution of the total magnetic field $B_{\text{tot}} = \sqrt{B_{x}^{2} + B_{y}^{2} + B_{z}^{2}}$ as a function of time for different merger scenarios with different initial magnetic fields, disc orientations and resolutions. Fig. 5a shows this evolution for the five merger scenarios with the smallest companion galaxies and with the standard initial magnetization of $10^{-9}$ G. Before the first encounter at $t = 1.3$ Gyr, there is an initial amplification of the galactic magnetic field similar to the lower-mass-ratio scenarios (cf. Fig. 4, but note the different time scales). An exception are the two merger scenarios with the smallest companion galaxies  M50 and M100, which show an even stronger initial amplification by approximately a factor of 8 due to the winding process. For these simulations with very small companion galaxies we used a factor of 10 higher resolution (see Table 2). The stronger amplification can be ascribed to the higher resolution (see chapter 4.2.4 and 4.4). 
Within the scenarios involving companion galaxies M10, M15 and M20, the amplification during the first encounter results in a galactic magnetic field strength lower than $10^{-6}$ G, which can be ascribed to the lower amount of impact energy available for conversion into turbulent energy (cf. chapter 4.3.2).  The maximum value of the galactic magnetic field strength ($1.9\cdot10^{-6}$ G, $1.0\cdot10^{-6}$ G and $0.7\cdot10^{-6}$ G for M1M10, M1M15 and M1M20 scenario, respectively) is reached at the time of the second encounter. Similar to the scenarios with smaller mass ratios (cf. Fig. 4), subsequent encounters and the final merger do not lead to any considerable further amplification of the magnetic field.

Generally, we find smaller saturation values of the galactic magnetic field for mass ratios of 10:1 and higher. Note, however, that the merging timescale increases with decreasing mass of the companion galaxy, wherefore we were not able to simulate the whole merging process for the M1M20 scenario and also the (high resolution) scenarios M1M50$\_$G9I9 and M1M100$\_$G9I9. Nevertheless, we can asses a general trend: Within the scenarios described above (e.g. the M1M10 and M1M15), the galactic magnetic field strength after the final merger does not exceed the maximum value reached after the first or second encounter. Therefore we do not expect a notable further amplification of the galactic magnetic field for the merger scenarios with mass ratios of 20:1 and higher.
As the high-mass-ratio scenarios M1M50$\_$G9I9 and M1M100$\_$G9I9 show only a slight amplification after the first encounter (with maximum values within the simulation time of $8.3\cdot10^{-8}$ G and $2.6\cdot10^{-8}$ G for M1M50 and M1M100 scenario, respectively), we expect the saturation value of the magnetic field to be also of order $10^{-8}$ G.
The final magnetic field strengths of the merger scenarios with mass ratios of 10:1 and lower saturate at values of order $10^{-6}$ G, whereas scenarios with mass ratios of 15:1 and higher saturate at lower values.

In summary, galaxy mergers with smaller companion galaxies show a slightly lower maximum value of the magnetic field strength and a flatter slope of the amplification. At the time of the final merger, the saturation values for scenarios with a mass ratio up to 10:1 show similar values of order $\mu$G, whereas the saturation value for larger mass ratios is decreasing with decreasing mass of the companion galaxy.

\subsubsection{Dependence on the initial magnetization}

Fig. 5b shows the total magnetic field as a function of time for the merger scenarios M1M4 and M1M7, with the highest, i.e. present-day magnetization of $B_{\text{gal,0}} = 10^{-6}$ G and $B_{\text{IGM,0}}=10^{-9}$ G. The corresponding simulations with the smallest magnetic field of $B_{\text{gal,0}} = 10^{-9}$ G and $B_{\text{IGM,0}}=10^{-12}$ G are shown for comparison (cf. Fig. 4c \& d). For the present-day magnetization we find only a small peak in the galactic magnetic field evolution at time of the first encounter, followed by a slight amplification. The maximum value of the galactic magnetic field strength ($5.3\cdot10^{-6}$ G and $3.8\cdot10^{-6}$ G for M1M4 and M1M7 scenario, respectively) is reached before the second encounter. This maximum value is reached at a earlier time and achieves a higher value compared to the scenarios with the lower initial magnetic field ($4.5\cdot10^{-6}$ G and $2.7\cdot10^{-6}$ G, respectively). Subsequently, the magnetic field in the present-day scenarios is 
decreasing. The saturation value within these scenarios is comparable to the final value within the simulations with lower initial magnetizations. Thus, the initial magnetic field strength of the scenarios has indeed an influence on the evolution of the galactic magnetic field. However, the saturation value of the galactic magnetic field is independent from the initial galactic magnetic field strength, which is consistent with previous studies and will be discussed in more detail in section 4.3.2.

\subsubsection{Dependence on the disc orientation}

Fig. 5c shows the magnetic field evolution as a function of time for the two retrograde scenarios M1M3 and M1M7 with the standard initial magnetization of $10^{-9}$ G (cf. Table 4) together with the corresponding prograde scenarios.

The galactic magnetic field amplification seems to be slightly more efficient for retrograde encounters than for prograde encounters. This can be ascribed to the presumably more efficient turbulence driving during the first encounter, as the different rotational velocities within the retrograde scenarios are expected to serve as additional shearing motions. Also, the maximum value of the magnetic field strength is higher for the retrograde scenario with the larger companion galaxy M3 ($7.2\cdot10^{-6}$ G), compared to the corresponding prograde scenario ($4.1\cdot10^{-6}$ G). The galactic magnetic field strength of the retrograde scenario M1M7 is smaller than the field strength in the corresponding prograde scenario until the second encounter. However, the saturation values of the galactic magnetic field within the retrograde scenarios are comparable to the values reached in the corresponding prograde scenarios. Thus, it is interesting to note that prograde or retrograde orientation of the galactic disc seems to influence the evolution of the galactic magnetic field during the collision, but shows only marginal effects on the saturation values of the galactic magnetic field.

\subsubsection{Dependence on the resolution}

Fig. 5d shows the magnetic field evolution as a function of time for the scenario M1M3 for the three different resolutions (cf. Table 4).
The general amplification behaviour does not change significantly with resolution. However, some features in the magnetic field evolution seem to be only resolved within the simulation with the highest resolution, particularly the remarkably steeper increase of the magnetic field strength after the first encounter. The simulation with the highest resolution shows a higher initial amplification of the galactic magnetic field prior to the first encounter. The final saturation value of the magnetic field is comparable for all resolutions, thus showing that a variation of the resolution does not alter the general results. This behaviour is due to the better resolution of turbulent structures in simulations with higher resolution, while the turbulent energy density remains the same for different resolutions.

\subsubsection{IGM magnetic field}

The IGM magnetic field saturates at a value of order $n$G within most of the scenarios. Thus, an initial magnetic field of $B_{\text{IGM,0}} = 10^{-9}$ G does not significantly grow during the interaction, whereas an initial field of $B_{\text{IGM,0}} = 10^{-12}$ G is clearly amplified up to the saturation value by three orders of magnitude. Within the simulations with an initial galactic magnetic field of $B_{\text{gal,0}} = 10^{-6}$ G, the IGM magnetic field amplification at the time of the first encounter is slightly more efficient because magnetic energy is additionally transported from the galaxies into the IGM by interaction-driven outflows \citep{KoLe11}. However, at the end of the simulations, we find the same saturation value of the IGM magnetic field as within all other scenarios. Moreover, the saturation value of the IGM magnetic field is on general independent on the masses of the progenitor galaxies up to a mass ratio of approximately 10:1. For larger mass ratios, the IGM magnetic field saturates at values slightly lower than order $n$G. This indicates that above a ratio of 10:1 the interaction is not violent enough to release a notable amount of energy into the IGM and thus to amplify the IGM magnetic field.

\subsection{Self-regulation of the amplification}

\begin{table}
\caption{Impact velocities and energies, maximum magnetic field strength reached between first and second encounter}
\begin{center}
\renewcommand{\arraystretch}{1.2}
\begin{tabular}{lllll}
\hline\hline
Collision&$v_{\text{Gal1}}$&$v_{\text{Gal2}}$&$E_{\text{impact}}$&$B_{\text{max}}$\\
&[km/s]&[km/s]&[10$^{42}$erg]&[10$^{-6}$G]\\\hline
M1M2&2.92&3.13&1.76&4.6\\
M1M3&2.71&3.44&1.49&4.1\\
M1M4&2.21&4.12&1.20&4.1\\
M1M5&1.95&4.52&1.05&3.4\\
M1M6&1.66&4.95&0.91&3.6\\
M1M7&1.43&5.28&0.80&2.2\\
M1M8&1.29&5.49&0.73&2.1\\
M1M9&1.18&5.74&0.68&1.8\\
M1M10&1.05&5.93&0.61&1.9\\
M1M15&0.84&6.45&0.47&1.0\\
M1M20&0.75&6.88&0.39&0.7\\
M1M50&0.65&7.38&0.20&-\\
M1M100&0.56&7.83&0.10&-\\
\hline
\end{tabular}
\end{center}
\label{2}
\end{table}

As we have shown in section 4.2., encounters of the galaxy M1 with a companion galaxy lead to a slightly higher maximum value of the total galactic magnetic field the lower the mass ratio of the progenitor discs and the larger the companion galaxy, respectively. This behaviour can be understood in terms of the impact energy of the colliding galaxies.

\begin{figure}
\vspace*{-6.7cm}
\hspace*{-1.8cm}
 \begin{minipage}{\textwidth}
	\begin{center}
\includegraphics[width=0.92\textwidth]{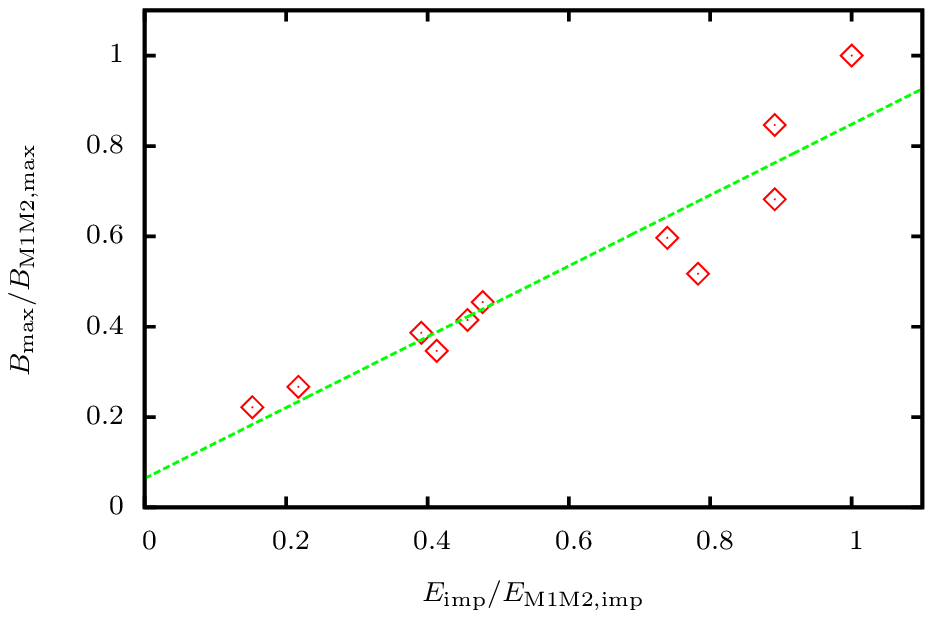}
	\end{center}
\end{minipage}
\vspace*{-12.cm}
\caption{\small{ Maximum magnetic field strength $B_{\text{max}}$ (reached between the first and second encounter within each scenario) as a function of the impact energy $E_{\text{imp}}$, each normalized to the values for the M1M2 merger scenario (cf. Table 5). There is a clear linear correlation (green line) between the the maximum magnetic field strength and the impact energy of the progenitor galaxies.}}
\end{figure}

\subsubsection{Impact energy}

\begin{figure}

 \begin{minipage}{\textwidth}

\includegraphics[width=8cm]{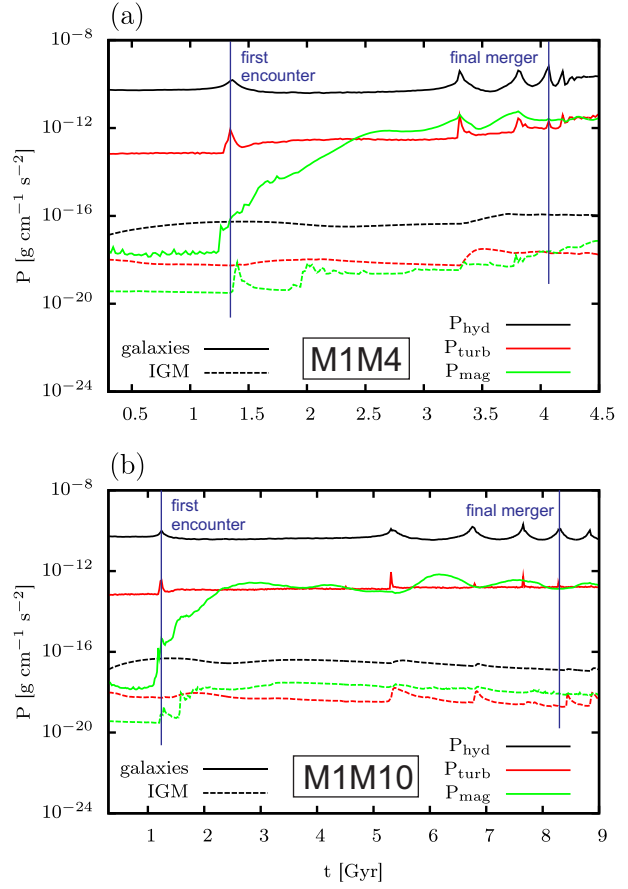}

\end{minipage}
\caption{\small{Evolution of the pressure components $P_{\text{hyd}}$ (black lines), $P_{\text{turb}}$ (red lines) and $P_{\text{mag}}$ (green lines) as a function of time for the scenarios M1M4 (a) and M1M10 (b) with the standard initial magnetization of $10^{-9}$ G. Pressure values for galaxies (solid lines) and IGM (dashed lines) are plotted separately using a density threshold of $10^{-29}$ g cm$^{-3}$. At the time of the final merger, the magnetic and the turbulent energy densities show approximately the same order of magnitude.}}
\end{figure}

The kinetic energy of the progenitor galaxies, which is released during the interaction, is expected to be partly converted into magnetic energy. Thus, the higher the impact energy, the higher the amount of turbulence and the more efficient the expected field amplification. 
The impact energy can be estimated on the basis of the masses and the center-of-mass-velocities of the progenitor galaxies just before the first encounter (this velocity is much less than the initial velocity of the galaxies since the galactic haloes already overlap, which causes a slowdown of the galaxies). The impact energy may be approximated by
\begin{equation}
E_{\text{impact}}=\frac{1}{2}m_{\text{G1}}v_{\text{G1}}^{2} + \frac{1}{2}m_{\text{G2}}v_{\text{G2}}^{2}.
\end{equation}
Center-of-mass-velocities and impact energies for the different merger scenarios are listed in Table 5.
As we set the galaxies on nearly-parabolic Keplerian two-body orbits, the center-of-mass velocity of the companion galaxy increases with decreasing mass, whereas the center-of-mass velocity of the larger galaxy M1 is slightly decreasing with decreasing mass of the companion galaxy. As a result, we find the impact energy of the M1M5 model to be roughly half the energy of the M1M2 model and the energy of the M1M10 model to be roughly one third of the impact energy of our largest merger model M1M2. This is consistent with the trend that scenarios with smaller companion galaxies show lower maximum values of the galactic magnetic field strength (cf. Table 5)  and also flatter slopes of the amplification during the interaction (cf. section 4.2.). The reason for this behaviour can most probably be ascribed to the lower impact energy which is available for conversion into magnetic energy. The resulting correlation between the maximum magnetic field strength and the impact energy is shown in Fig. 6.

\begin{figure*}

 \begin{minipage}{\textwidth}
	\begin{center}
\includegraphics[width=\textwidth]{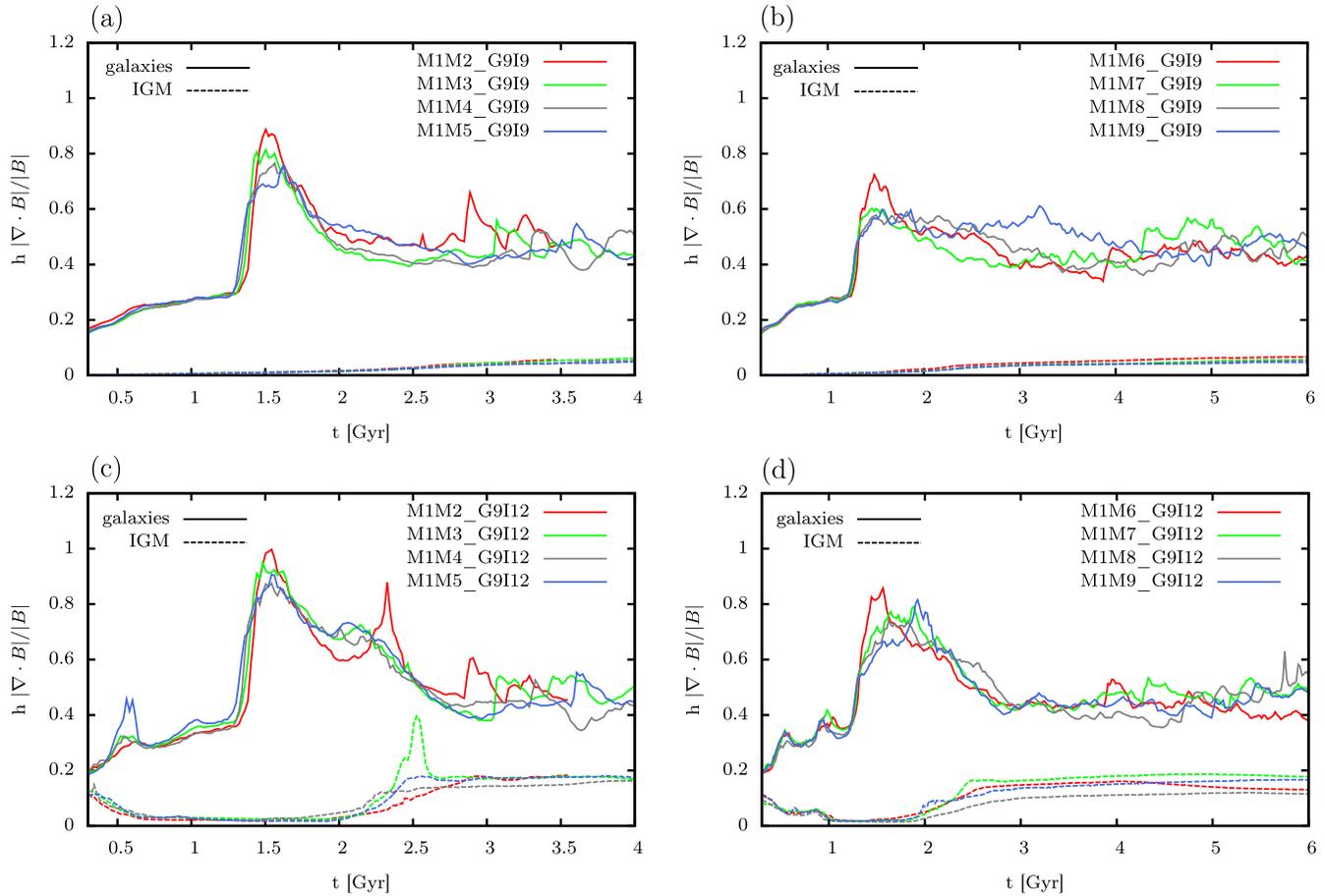}
	\end{center}

\caption{\small{ Mean numerical divergence measure $\langle h|\nabla\cdot\textbf{B}|/|\textbf{B}|\rangle$ as a function of time for eight collision simulations with the standard initial magnetization of $10^{-9}$ G (\textit{upper panels}) and a lower initial IGM magnetic field of $B_{\text{IGM},0}=10^{-12}$ G (\textit{lower panels}). Galaxies (solid lines) and IGM (dashed lines) are shown separately using a density threshold of $10^{-29}$ g cm$^{-3}$. The numerical divergence measure stays below the tolerance value of unity during all of the simulations.}}
\end{minipage}
\label{div}
\end{figure*}

\subsubsection{Pressures}
The release of kinetic energy during the interaction drives turbulence, which in turn results in an amplification of the magnetic field by magnetic field line folding and shearing \citep[see e.g.][for a review]{BrSu05}. However, if the magnetic energy density reaches the magnitude of the turbulent energy density, the magnetic field amplification caused by the turbulent motion of the gas is suppressed by the magnetic field itself via the Lorentz force. The system therefore tends to maintain a dynamic equilibrium or equipartition between turbulent and magnetic energy density \citep[see e.g.][]{Be07,ChBo07,ArBe09,KoKa10,KoLe11}.

In order to study the expected energy equipartition between the magnetic and the turbulent energy density (or, equivalently, magnetic and turbulent pressure) in more detail, we follow \citet{KoLe11} and choose $v_{\text{rms}}$ (rms velocity around the mean velocity inside the smoothing length $h$) as an estimator for the local turbulent velocity\footnote{$v_{\text{rms}}$ is a comparatively conservative estimator for the turbulent velocity \citep[see][]{KoLe11}. Hence, it cannot be ruled out that the turbulent velocity may be overestimated on small-scales and underestimated on larger scales.}. The turbulent pressure is then given by $P_{\text{turb}}=1/2 \rho v_{\text{rms}}^{2}$. 

Fig. 7 shows the turbulent pressure, the hydrodynamic pressure $P_{\text{hyd}}=1/2 \rho v^{2}$ (with $v$ the total velocity of each particle), and the magnetic pressure $P_{\text{mag}}=B^{2}/8\pi$ exemplarily for the M1M4 and the M1M10 scenarios with the standard magnetization of $10^{-9}$ G. The pressure values for the galaxies (solid lines) and the IGM (dashed lines) are plotted separately using a density threshold of $10^{-29}$g cm$^{-3}$. 
While the hydrodynamic pressure within the galaxies in both scenarios evolves relatively smoothly and stays in the same range of magnitude during the whole interaction (except for the peaks indicating the different encounters of the discs), the turbulent pressure slightly increases after each encounter. During the collision, the galactic turbulent pressure lies below the galactic hydrodynamic pressure by roughly two orders of magnitude. The magnetic pressure gets strongly amplified during the interaction (according to the amplification of the magnetic field itself), as has been already shown for major galaxy collisions by \citet{KoKa10,KoLe11}. 

Within both scenarios, the system reaches the equipartition level between turbulent and magnetic pressure at about 1 Gyr after the first encounter. In case of the larger companion galaxy in scenario M1M4 (Fig. 7a), the magnetic pressure stays slightly above the turbulent pressure until the final merger, whereupon an approximate equipartition is reached. During the longer period between first and second encounter of the scenario M1M10 (Fig. 7b), the  magnetic pressure decreases again due to the dilatation of the shocked region, which causes also a dilatation of the magnetic field energy. However, in the subsequent evolution the magnetic pressure gets amplified again with each further encounter, whereas each amplification period is followed by a period of decreasing magnetic field until the next encounter takes place. At the time of the final merger, the magnetic and the turbulent energy densities show approximately the same order of magnitude. As the equipartition value within each model depends on the energy supplied to the system (impact energy), we generally find higher equipartition values for lower mass ratios of the galaxies.

Within each of the scenarios shown in Fig. 7, the IGM shows a slight increase in the hydrodynamic pressure at the beginning of the simulation, followed by a relatively smooth evolution. The IGM turbulent pressure clearly shows some peaks, corresponding to the encounters. The IGM turbulent pressure lies below the IGM hydrodynamic pressure by roughly two orders of magnitude during the whole simulation, which is a comparable separation as for the corresponding galactic pressure components. The IGM magnetic pressure within each scenario gets amplified during the first encounter. In the subsequent evolution, a loose equipartition between the IGM turbulent and magnetic pressure is maintained until the end of the simulations. 

In summary, within all of our merger simulations, we see a general trend: magnetic and turbulent pressures of galaxies and IGM tend to reach equipartition, which is consistent with studies of major galaxy collisions \citep{KoKa10,KoLe11} and in good agreement with theoretical expectations \citep{ArBe09}.

\subsection{Numerical divergence and resolution study}
In SPMHD simulations the numerical divergence measure $\langle h|\nabla\cdot\textbf{B}|/|\textbf{B}|\rangle$ is generally 
regarded as a measure of reliability of the simulations. It is important to note that the numerical divergence is not a physical divergence caused by magnetic monopoles. \citet{KoLe09} have demonstated that the numerical divergence in simulations of isolated galaxies performed with Euler Potentials (which are free of a physical divergence by construction) can still reach values up to order of unity. Hence, the value of the numerical divergence measure in SPMHD simulations with direct magnetic field implementation should not exceed unity in order to guarantee the reliability of the simulation results. 
As already pointed out by \citet{KoKa10} and \citet{BuCl11}, the SPH estimator for the divergence $(\nabla\cdot\textbf{B})_{i}$ (with i being the index of the particle) accounts for weighted magnetic field differences of neighbouring particles within the smoothing length $h_{i}$, and is
 therefore primarily dependent on the irregularity of the magnetic field within the kernel. In order to avoid instabilities due to the numerical divergence errors, a subtraction of any non-vanishing divergence from the momentum equation (see chapter 2.2) is adopted for the simulations presented in this paper. Also, the adopted IGM helps to increase the numerical stability, as the magnetic field is not dropping to zero at the edges of the galaxies, which could lead to incorrect calculations of the numerical divergence in these regions.

Fig. 8 shows the mean numerical divergence measure $\langle h|\nabla\cdot\textbf{B}|/|\textbf{B}|\rangle$ as a function of time for eight different merger scenarios with the standard initial magnetization of $10^{-9}$ G (Fig. 8a \& b) and a lower initial IGM magnetic field of $B_{\text{IGM,0}} = 10^{-12}$ G (Fig. 8c \& d).
The values of the numerical divergence stay below unity within all of the simulations and approach unity only during the first encounter. On general, the divergence measure is higher for smaller initial magnetizations, in agreement with previous studies. This behaviour follows from the fact that in case of weaker magnetic fields the encounters exert stronger effects on the irregular motions of the particles, which in turn are responsible for the numerical divergence \citep[cf.][]{KoKa10}.

Fig. 9 shows the mean numerical divergence measure $\langle h|\nabla\cdot\textbf{B}|/|\textbf{B}|\rangle$ as a function of time for the three different resolutions of scenario M1M3 with the standard magnetization of $10^{-9}$ G. For the high resolution (green lines), the numerical divergence measure is by a factor of two lower than for the medium resolution (blue lines), and for the low resolution (red lines), it is by a factor of two higher than for the medium resolution. This can be explained considering the smoothing length $h_{i}$, which is smaller for a higher resolution. For a smaller smoothing length, we expect less sub-resolution tangling of the magnetic field, resulting in a lower numerical divergence measure for a higher resolution.

\begin{figure}
 \begin{minipage}{\textwidth}
\includegraphics[width=.45\textwidth]{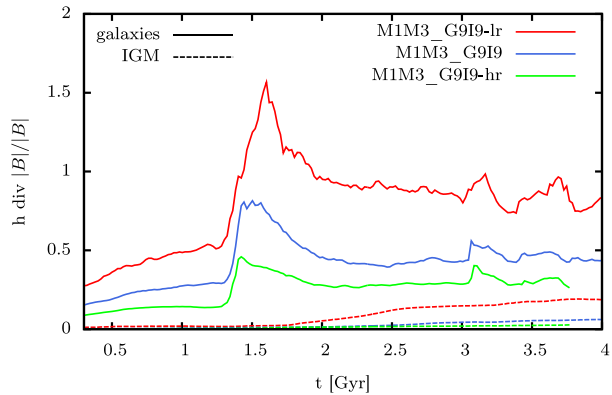}
\end{minipage}
\label{res}
\caption{\small{ Mean numerical divergence measure $\langle h|\nabla\cdot\textbf{B}|/|\textbf{B}|\rangle$ as a function of time for the M1M3 scenario with the standard magnetization of $10^{-9}$ G, using three different resolutions. Galaxies (solid lines) and IGM (dashed lines) are shown separately using a density threshold of $10^{-29}$ g cm$^{-3}$. The divergence measure is higher for lower resolutions.}}
\end{figure}

On the other hand, the saturation value of the magnetic field does on general not depend on the resolution (Fig. 5d), showing that even a by a factor of two higher numerical divergence would not alter the general results. Hence, we conclude that the presented medium-resolution simulations are numerically reliable.

\begin{figure*}
 \begin{minipage}{\textwidth}
	\begin{center}
\includegraphics[width=16cm]{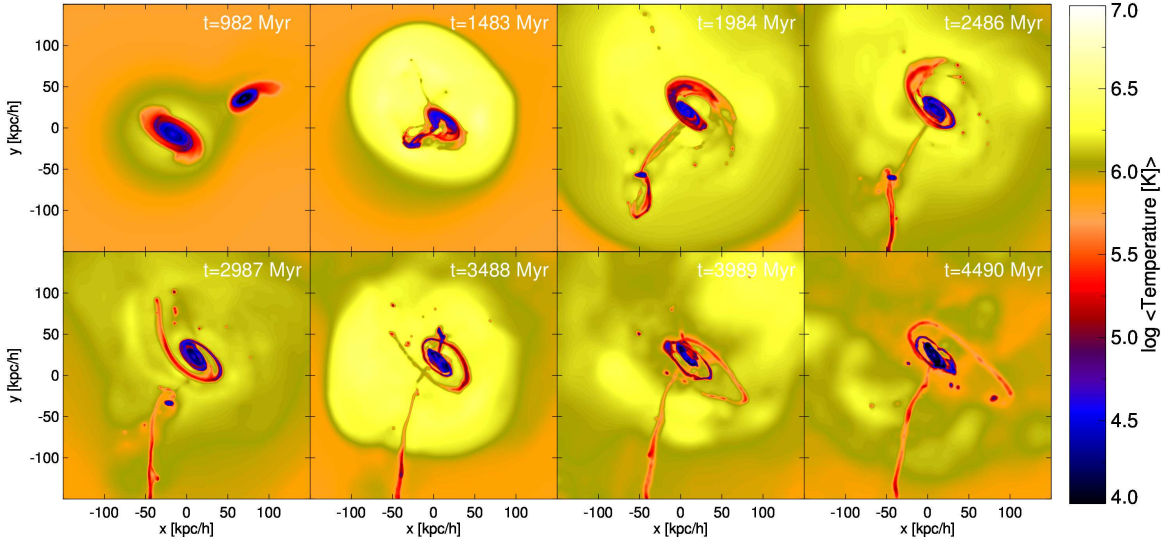}
	\end{center}
\end{minipage}
\caption{\small{ Evolution of the projection of the mean line-of-sight temperature at the same time steps and for the same scenario as in Fig. 1.}}
\label{temp1}
\end{figure*}

\subsection{Temperature evolution and X-ray emission}
The X-ray emission during a galactic encounter is expected to depend on the orbit, the disc orientation, the masses and the gas fraction of the progenitor discs \citep[e.g.][]{CoDi06}. 
The main source of X-ray emission is hot gas, which is heated by shocks accompanying the interaction. As the gas density within the IGM and the strength of the magnetic field have an impact on the gas behaviour during a galactic interaction, they may be expected to also alter the X-ray emission by a non-negligible order of magnitude. 

As shown in \citet{KoLe11}, the propagation velocity and thus the Mach numbers of shocks driven by galactic collisions are higher for stronger initial magnetic fields. This is due to the magnetic pressure, which is supporting the shocks and galactic outflows. Also, the higher this additional pressure component, the less the gas can compress, thus leading to lower gas densities. Higher Mach numbers are in turn related to the temperature of the IGM behind the shock fronts (Rankine-Hugoniot shock jump conditions). Additionally, the region of gas which is heated by the shock is larger for a higher shock propagation velocity and a stronger galactic outflow, respectively. As a consequence, the X-ray emission may be expected to increase with increasing magnetic field strength \citep[cf.][]{KoLe11}.

We want to investigate the dependence of the X-ray emission on the magnetic field strength in the presence of an ambient IGM \citep[thus expanding previous studies, e.g.][]{CoDi06,SiHo09}. Therefore, we have calculated the bolometric X-ray luminosity following the method of \citet{NaFr95}. This method is based on the assumption that the main X-ray mechanism is thermal bremsstrahlung, in agreement with the applied zero-metallicity cooling function (cf. section 2.3). The bolometric X-ray luminosity is calculated according to \citep{NaFr95}
\begin{equation}
L_{\text{x, bolo}} = 1.2 \cdot 10^{-24} \frac{1}{\left(\mu m_{\text{p}}\right)^{2}} \hspace{1mm} \sum_{i=1}^{N_{\text{gas}}} m_{\text{gas},i} \hspace{1mm} \rho_{i} \left( \frac{k_{\text{B}} T_{i}}{\text{keV}}\right)^{1/2},
\end{equation}
with mass $m_{\text{gas},i}$, density $\rho_{i}$ and temperature $T_{i}$ of the $i$-th gas particle in cgs-units, respectively. Only fully ionized particles should be considered when calculating the luminosity. Therefore, we exclude contributions of particles with temperatures lower than $10^{5.2}$ K and densities higher than $0.01 \text{M}_{\odot} \text{pc}^{-3}$ \citep[cf.][]{CoDi06}.

\begin{table}
\caption{Maximum and final values of IGM and total temperature and total X-ray luminosity corresponding to Fig. \ref{LT}}
\begin{center}
\renewcommand{\arraystretch}{1.2}
\begin{tabular}{lllll}
\\ \hline\hline
Scenario&T$_{\text{max}}$&T$_{\text{final}}$&Lx$_{\text{max}}$&Lx$_{\text{final}}$\\
&[$10^{5}$ K]&[$10^{5}$ K]&[erg s$^{-1}$]&[erg s$^{-1}$]\\\hline
\multicolumn{5}{c}{\textsc{IGM}} \\ \hline
M1M4$\_$G6I9&7.26&4.74&38.57&36.56\\
M1M4$\_$G9I9&6.84&4.29&38.52&36.78\\
M1M4$\_$G0I0&5.88&4.56&38.49&37.51\\
M1M7$\_$G6I9&6.79&3.99&38.55&36.00\\
M1M7$\_$G9I9&6.12&4.61&38.50&36.42\\
M1M7$\_$G0I0&5.62&4.52&38.46&37.44\\\hline
\multicolumn{5}{c}{\textsc{total}} \\ \hline
M1M4$\_$G6I9&7.58&4.95&38.85&37.04\\
M1M4$\_$G9I9&6.86&4.53&38.97&37.33\\
M1M4$\_$G0I0&6.72&5.53&38.89&38.33\\
M1M7$\_$G6I9&6.99&4.12&38.78&36.68\\
M1M7$\_$G9I9&6.49&4.69&38.74&36.73\\
M1M7$\_$G0I0&6.36&5.40&38.80&38.15\\
\hline
\end{tabular}
\end{center}
\end{table}

\begin{figure*}
 \begin{minipage}{\textwidth}
\vspace*{1.0cm}
	\begin{center}
\hspace*{-.4cm}
\includegraphics[width=1.\textwidth]{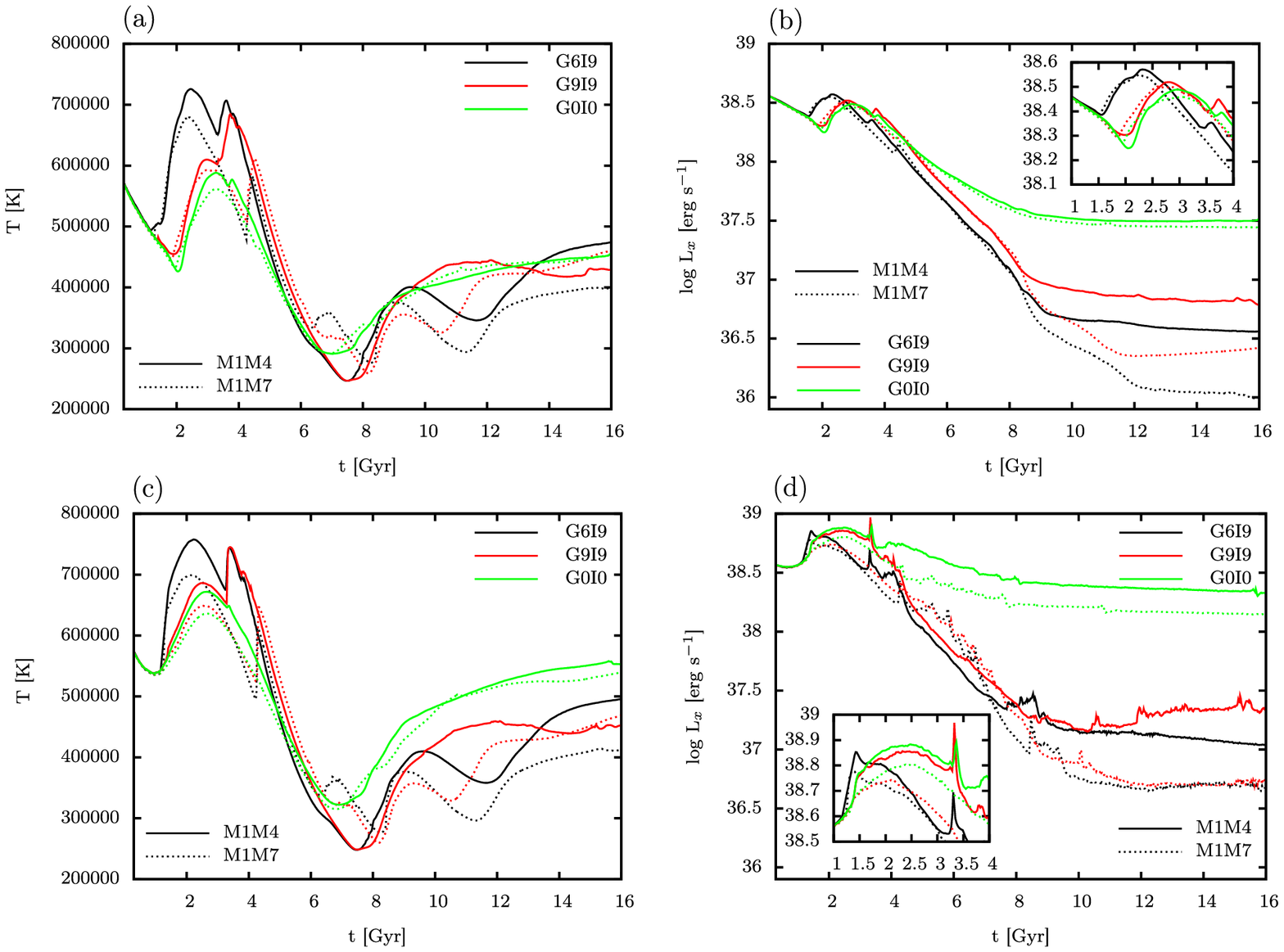}
	\end{center}
\vspace*{1.3cm}
\end{minipage}
\caption{\small{Mean temperature (left panels a\&c) and mean bolometric X-ray luminosity (calculated according to Eq. (17), right panels b\&d) of the fully ionized particles as a function of time for the merger scenarios M1M4 (\textit{solid lines}) and M1M7 (\textit{dotted lines}), and with different initial magnetizations of  $B_{\text{gal,0}}=10^{-6}$ G, $B_{\text{IGM,0}}=10^{-9}$G (black lines), $B_{\text{gal,0}}= B_{\text{IGM,0}}=10^{-9}$G (red lines) and excluding magnetic fields (green lines). The upper panels show the temperature and luminosity of the IGM (applying the same threshold as in Fig. 4-7), and the lower panels show the total (galaxies plus IGM) quantities. The higher the initial magnetization, the more efficient the increase of the temperature and thus the X-ray luminosity within the IGM after the first encounter at $t \approx 1.3$ Gyr. However, the presence of magnetic fields does not enhance the total X-ray luminosity. After the first encounter, the total X-ray luminosities decrease faster and result in a lower final value of the luminosity for higher initial magnetizations. This may be explained by the stronger interaction-driven outflows within the simulations with higher initial magnetizations, which lead to a faster dilution of the IGM gas and thus a lower gas density within the X-ray emitting regions.}}
\label{LT}
\end{figure*}

\subsubsection{Temperature evolution}
Fig. \ref{temp1} shows the morphological evolution of the projection of the mean line-of-sight temperature for the present-day scenario M1M4\_G6I9, whereby the eight panels display a sequence of increasing time. Before the first encounter at about 1.3 Gyr, the galaxies are surrounded by a small region of hot gas due to the initial accretion of IGM gas onto the galaxies, whereas the gas within the galaxies is cooler (Fig. \ref{temp1}, left upper panel). During the first encounter, shocks and outflows are driven into the IGM, thus heating the IGM gas (Fig. \ref{temp1}, second upper panel) \citep[cf. also][]{KoLe11}. After the first encounter, the temperature of the gas surrounding the galaxies decreases again due to the dilatation of the shocked IGM (Fig. \ref{temp1}, left lower panel). Subsequent encounters result in a further shock-heating of the IGM gas. After the final merger ($\approx$ 4 Gyr), the IGM gas slowly cools down again.

Fig. \ref{LT} shows the temperature (left panels) and the bolometric X-ray luminosity (right panels) of the fully ionized particles, respectively, as a function of time for the merger scenarios M1M4 (solid lines) and M1M7 (dotted lines),  and with different initial magnetizations of $B_{\text{gal,0}}=10^{-6}$ G, $B_{\text{IGM,0}}=10^{-9}$G (black lines), $B_{\text{gal,0}}= B_{\text{IGM,0}}=10^{-9}$G (red lines) and excluding magnetic fields (green lines). The upper panels show the temperature and luminosity of the IGM, and the lower panels show the total (galaxies plus IGM) quantities. Table 6 summarizes the maximum temperatures (T$_{\text{max}}$) and X-ray luminosities (Lx$_{\text{max}}$) reached during the interaction, as well as the final values (T$_{\text{final}}$ and Lx$_{\text{final}}$) reached at $t = 16$ Gyr (cf. Fig. \ref{LT}).

\begin{figure*}
 \begin{minipage}{\textwidth}
	\begin{center}
\includegraphics[width=\textwidth]{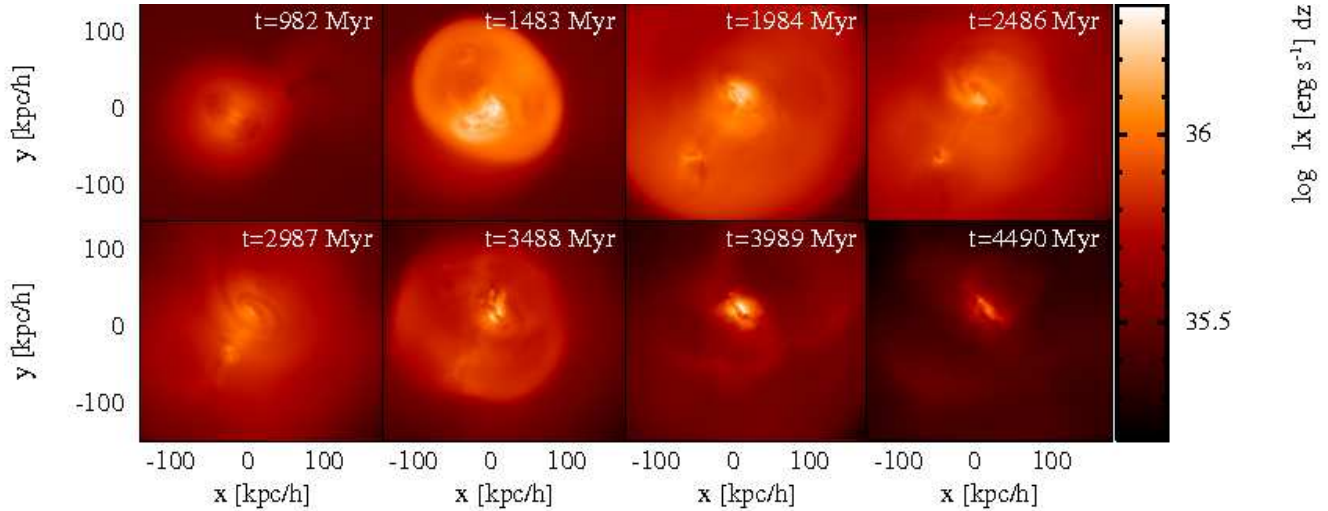}
	\end{center}
\vspace*{-6.8cm}
\end{minipage}
\caption{\small{ Evolution of the projection of the mean line-of-sight X-ray luminosity at the same time steps and for the same scenario as in Fig. 1.}}
\label{X1}
\end{figure*}

Fig. \ref{LT}a shows the mean IGM temperature as a function of time, applying the same density threshold as before in Figs. 4-7. 
We have set the initial IGM temperature to the virial temperature of the larger galaxy M1 (cf. section 3.4), thus overestimating the equilibrium temperature of the merging system. Thus, the IGM gas tends to cool in the very beginning of the simulations. At the time of the first encounter ($t\approx 1.3$ Gyr), interaction-driven shocks reheat the IGM gas, resulting in temperatures considerably higher than the initial temperature. Thereby, the increase of the IGM temperature is more efficient for stronger initial magnetizations, in agreement with the previous studies \citep[cf.][]{KoLe11}, reaching higher maximum temperatures of the IGM for higher initial magnetic field strengths (cf. Table 6). Furthermore, the maximum temperature is higher for larger companion galaxies, i.e. for higher impact energies (cf. Table 5). After the final merger ($t \approx 4$ Gyr for the M1M4 and $t \approx 6.5$ Gyr for the M1M7 scenario) the IGM temperature decreases. The final IGM temperatures lie between $4\cdot10^{5}$ K and $4.7\cdot10^{5}$ K within all of the scenarios shown in Fig. 11 (see also Table 6).

Fig. \ref{LT}c shows the mean total temperature as a function of time. The behaviour of the total temperature is generally very similar to the behaviour of the IGM temperature, because the interaction-driven shocks propagate mainly into the IGM and deposit their energy there. The maximum total temperatures reached during the first encounter are slightly higher than the  maximum IGM temperatures. However, the spread in the final total temperatures is considerably larger compared to the IGM values (see Table 6).

\subsubsection{Evolution of bolometric X-ray emission}

Fig. \ref{X1} shows the morphological evolution of the projection of the mean line-of-sight X-ray emission for the present-day scenario M1M4\_G6I9, whereby the eight panels display the same sequence of increasing time as in Fig. 1. Before the first encounter (upper left panel), the X-ray luminosity is relatively high within the galaxies (high densities) and the surrounding region (higher temperatures due to the initial accretion of the IGM gas, cf. the previous section). Until the final merger ($t \approx 4$ Gyr for the M1M4 and $t \approx 6.5$ Gyr for the M1M7 scenario), the X-ray luminosity evolves according to the temperature evolution, i.e. decreases until the first encounter due to the initial cooling of the IGM gas, increases during the first encounter due to the shock-heating of the IGM, and subsequently decreases due to gas cooling and dilatation of the shock-heated regions.

Fig. \ref{LT}b shows the bolometric luminosity as a function of time for the IGM. The correlation between luminosity and temperature (Fig. 11a) until the final merger is clearly visible. Particularly, the luminosity increases considerably during the first and second encounters ($t_{\text{first}} \approx 1.3$ Gyr and $t_{\text{second}} \approx 3.3$ Gyr for the M1M4 and $t_{\text{second}} \approx 4.3$ Gyr for the M1M7 scenario, respectively, inlay in Fig. 10b), whereby higher initial magnetic field strengths result in a faster and stronger increase of the IGM X-ray luminosity directly after the first encounter and also higher maximum values. However, after the final merger, the IGM X-ray luminosity decreases even more than the IGM temperature. This is due to the linear dependence of the luminosity on the density (cf. Eq. 17) and the dilution of the shock-heated IGM gas (not shown).
The final values of the IGM X-ray luminosity lie in the range between $36.0$ erg s$^{-1}$ and $37.5$ erg s$^{-1}$ for all of the merger scenarios. Most interestingly, the final luminosities are higher for \textit{lower} initial magnetizations. Additionally, they are also higher for larger companion galaxies, i.e. higher impact energies (see Table 5). We will discuss this behaviour in more detail in the next section.

Fig. \ref{LT}d shows the mean total bolometric luminosity of both the IGM and the galaxies as a function of time. As within the IGM, the behaviour of the total X-ray luminosity follows the behaviour of the total temperature until the final merger (thus also reaching higher maximum X-ray values compared to the IGM maximum values) and subsequently decreases faster than the temperature due to the additional dilatation of the IGM gas. The final values of the total X-ray luminosity ($t \approx 16$ Gyr) lie in the range between $36.7$ erg s$^{-1}$ and $38.3$ erg s$^{-1}$, i.e. are slightly higher as for the IGM only. As before, the final luminosities are higher for lower magnetizations and larger companion galaxies, i.e. higher impact energies.

\subsubsection{Discussion of X-ray emission}

The presence of magnetic fields shows a clear influence on the evolution of the total X-ray luminosity in the presented simulations (Fig. 11). On the one hand, interaction-driven shocks have been shown to gain higher Mach numbers in the presence of a strong magnetic field \citep{KoLe11}, thus leading to a more efficient shock-heating of the IGM gas and consequently a higher X-ray luminosity within the IGM. In this sense, this is a positive feedback of the magnetic field on the X-ray luminosity. However, the X-ray emission of the galaxies themselves outweights the emission of the IGM at the times of the different encounters (see the inlays in Fig. 11 b \& d, particularly the clear peaks in Fig. 11d). On the other hand, there is also a negative magnetic feedback: The stronger the magnetization of the system, the higher the magnetic pressure pushing the gas apart, and thus the more efficient the dilution of the shock-heated gas. Given that the X-ray emission scales with the root of the temperature but squared with the gas density (Eq. 17), this negative feedback soon overtakes the positive, finally resulting in a clearly lower final X-ray luminosity of the system in the presence of stronger magnetic fields (Fig. 11 and Table 6). This is contrary to what was expected by \citet{KoLe11}.

In general, we find relatively low final values of the total X-ray luminosity ($38.15-38.33$ erg s$^{-1}$ in our simulations without magnetic fields) compared to the final X-ray luminosity values found in \citet{SiHo09} ($\approx 38.5-41.5$ erg s$^{-1}$, depending on the gas content and the mass of the progenitor galaxies, cf. their Fig. 6) and \citet{CoDi06} ($\approx 39-42.5$ erg s$^{-1}$, depending on the progenitor mass, cf. their Fig. 11). This is most probably due to the lack of an accreting black hole in our models, and that zero-metallicity cooling is used for the X-ray computation in contrast to the metal-line cooling\footnote{Because our intention is to investigate the influence of magnetic fields on $L_{\text{x, bolo}}$, we are focusing on the more conservative zero-metallicity method. As shown by \citet{CoDi06}, the progression of the X-ray emission is very similar for both methods, but the metallicity-dependent cooling leads to a higher X-ray luminosity during the collisions (see their Fig. 2). Thus, the zero-metallicity X-ray luminosity may be regarded as a lower limit \citep{CoDi06}.} applied by \citet{CoDi06}. Furthermore, as the overall X-ray emission depends also on the gas content, the orbit and the disc orientation \citep[cf.][]{CoDi06}, the initial setup of the scenarios has an additional influence on the X-ray luminosity. 

\citet{SiHo09} simulated interacting galaxies including hot gas within the galactic haloes, but no gaseous disc. This is a clear difference in the setup compared to our gaseous discs. However, despite this difference, we find the X-ray luminosity to increase and reach its maximum prior to the final merger, similar to what was found by \citet{SiHo09}. Furthermore, this evolution of the X-ray emission corresponds also to observational results, which show the X-ray luminosity to increase during the merger and peak about 300 Myr before the final merger \citep{BrPo07}.

\citet{CoDi06} considered the X-ray emission generated during interactions of galaxies including gaseous discs, but without surrounding IGM gas. In good agreement with their results, we generally find a higher X-ray luminosity for larger progenitor galaxies (cf. our Fig. 11 and their Fig. 11), due to the aspired hydrostatic equilibrium after the final merger. On the other hand, we find the maximum X-ray luminosity after the first or the second encounter, not at the time of the final merger as found by \citet{CoDi06}. 

Furthermore, our minor merger simulations show much longer time periods between the encounters, thus enabling cooling of the gas. This is different to the major merger studies with relatively short but violent cooling periods found by \citet{CoDi06}. Altogether, we find at least one order of magnitude lower X-ray luminosities compared to \citet{CoDi06} and a faster decrease in luminosity after the final merger. Again, this difference may result from the lack of black holes and metal-line cooling in our simulations, and the lack of IGM gas in the studies by \citet{CoDi06}.

In summary, we may underestimate the total X-ray luminosity within our simulations, but our results generally agree well with previous studies and observations. Therefore, we are confident that our conclusions on the impact of the magnetic field on the ability of a system to emit in X-ray are reliable.

\section{Conclusion and discussion}
We have presented a series of 32 galaxy minor mergers including galactic magnetic fields and a magnetized ambient IGM. We have investigated the evolution of the galactic and IGM magnetic field with respect to different mass ratios, initial magnetic field strengths, disc orientations and numerical resolutions. The main results can be summarized as follows:
\begin{itemize}
\item The smaller the companion galaxy, i.e. higher the mass ratio, the lower the maximum galactic magnetic field and the flatter the slope of the amplification. The magnetic field for mass ratios up to 10:1 saturates at values of order $\mu$G, whereas scenarios with smaller companion galaxies saturate at values lower than $\mu$G, suggesting that the impact energies for mass ratios above 10:1 are not high enough to provide enough energy for conversion into magnetic energy.
\item The saturation value of the galactic magnetic fields is independent on the initial galactic magnetic field strength, in good agreement with previous studies \citep{KoKa10,KoLe11}.
\item The disc orientation (prograde or retrograde) slightly changes the evolution of the galactic magnetic field during the encounter, but has only a minor effect on the saturation value of the galactic magnetic field.
\item Some features of the magnetic field evolution seem to be only resolved in simulation with the highest numerical resolution. However, the saturation value of the magnetic field is comparable for the three investigated resolutions. Most probably, this is due to the better treatment of small-scale turbulence in high-resolution simulations, which, however, does not significantly increase the total turbulent energy of the system.
\item The IGM magnetic field saturates at a value of order $10^{-9}$G within most of the scenarios. For mass ratios higher than 10:1, the IGM saturation value is slightly lower, which can again be traced back to the lower impact energy of these interactions.
\end{itemize}
Summing up, minor mergers up to a mass ratio of 10:1 are able to amplify a small ($10^{-9}$ G) galactic magnetic field up to $\mu$G order, a value which is in good agreement with observations \citep[see e.g.][]{BeBr96,ChBo07,VoSo10}. Also, the saturation value of the IGM magnetic field of $\approx 10^{-9}$ G is in very good agreement with the upper limit estimates of $10^{-9}-10^{-8}$ G derived from observations \citep{KrBe08}.

Furthermore, the maximum values of the galactic magnetic field reached during the interactions are higher and the slope of the amplification is steeper for lower mass ratios. This is reasonable because the main source for the magnetic field amplification is the impact energy released during the interaction. This impact energy is in turn higher for lower mass ratios (section 4.3.1 and Table 5).

Moreover, we find that the magnetic pressures associated with the IGM and galactic magnetic fields saturate at the equipartition level between turbulent and magnetic pressure (section 4.3.2), which is consistent with studies of major galaxy collisions \citep{KoKa10,KoLe11} and theoretical considerations. As the energy equipartition within each scenario depends on the impact energy supplied to the system, the equipartition values are on general higher for larger companion galaxies, i.e. lower mass ratios.

The values of the numerical divergence measure $h|\nabla\cdot\textbf{B}|/|\textbf{B}|$ (which are generally non-zero within SPMHD simulations) have been shown to lie below the tolerance-threshold of unity within all of our simulations (section 4.4). Thus, our simulations are numerically reliable \citep[cf.][]{KoKa10}. Moreover, our resolution study showed that the numerical divergence measure is lower for a higher numerical resolution, but simultaneously the saturation value of the magnetic field is generally independent of the resolution (section 4.2.4). Thus, our conclusions are robust with respect to the numerical divergence and resolution.

Finally, we presented a detailed analysis of the evolution of the temperature and the X-ray luminosity during the simulated minor merger scenarios (section 4.5). We find that a magnetic field has both, a positive and a negative feedback on the X-ray luminosity. On the one hand, a higher magnetic field results in higher Mach-numbers of the interaction-driven shocks and thus a more efficient shock-heating of the IGM \citep[cf.][]{KoLe11}. Consequently, the X-ray luminosity is higher within recently shock-heated regions. On the other hand, a higher magnetic pressure leads also to a more efficient dilution of the shock-heated regions, resulting in lower gas densites and hence lower X-ray luminosities. Given the different dependence of the X-ray luminosity on the temperature and the gas density (Eq. 17), the negative feedbeck is more important in the long run. Consequently, we find lower final X-ray luminosities for higher initial magnetizations, in contrast to what was expected by \citet{KoLe11}.

As galaxy mergers are believed to be an essential part of hierarchical growth of structure in the universe, these events may be expected to provide a non-negligible contribution to the amplification of magnetic fields on galactic scales and, given the simultaneous magnetization of the ambient IGM, also beyond. Commonly, the presence of galactic magnetic fields is explained by the action of the galactic dynamo \citep[e.g.][]{Fe99,GrEl08,HaOt09}. However, this hypothesis is challenged by recent observations of strong ($\mu$G) magnetic fields at high redshifts \citep{BeMi08}, because the galactic dynamo can amplify magnetic fields only on a timescale of several Gyr. Also, it has been shown that that dynamos cannot work efficiently within dwarf galaxies due to the lack of global differential rotation \citep{GrEl08}. 
However, \citet{DuTe10} presented a possible machanism for the magnetization of the early universe. Within their \textquotedblleft Cosmic Dynamo\textquotedblright \hspace{1mm} the universe is magnetized by the combined action of gravitational instabilities forming dwarf galaxies, a galactic dynamo working within these newly formed dwarf galaxies, and galactic winds expelled into the IGM at times of violent SF activity.

Nevertheless, galactic interactions provide a promising alternative or complement to the scenarios proposed so far: As galaxy mergers, and especially minor mergers, are believed to have been much more frequent in earlier times of the universe, it is likely that the presented interaction-driven amplification also provided for a significant contribution to the amplification of the galactic and intergalactic magnetic fields on short timescales. Hence, for future studies it would be interesting to focus on mergers of dwarf galaxies to gain a more complete picture of the evolution of magnetic fields within mergers. Furthermore, studies of the magnetic field evolution in the early universe in a cosmological context and within the formation of galaxies would be very worthwhile for a better understanding of the overall process of the magnetic field amplification caused by galaxy formation and interaction events in the history of the universe.

\section*{Acknowledgments}
A.G. thanks Volker Springel for the programs to set up the initial galaxy models.
Rendered plots were made using the \textsc{splash} software written by Daniel Price \citep[see][]{Pr07}, and with \textsc{smac} \citep[see][]{DoHa05} with contributions by Julius Donnert. Granting of computing time from John von Neumann-Institute for Computing (NIC), J\"{u}lich, Germany, is gratefully acknowledged.

K.D. acknowledges the support by the DFG Priority Programme 1177 and additional support by the DFG Cluster of Excellence \textquoteleft Origin and Structure of the Universe\textquoteright .

\end{document}